\documentclass[aps,amsmath,amssymb,twocolumn,prb,superscriptaddress]{revtex4-2}

\usepackage{graphicx}
\usepackage{dcolumn}
\usepackage{bm}
\usepackage{color}
\usepackage[normalem]{ulem}
\usepackage{amsmath}
\usepackage{hyperref}

\begin{document}

\title{Sextets in four-terminal Josephson junctions}

\author{Miriam Rike Ebert}
\affiliation{Fachbereich Physik, Universit{\"a}t Konstanz, D-78457 Konstanz, Germany}

\author{David Christian Ohnmacht}
\email{david.ohnmacht@uni-konstanz.de}
\affiliation{Fachbereich Physik, Universit{\"a}t Konstanz, D-78457 Konstanz, Germany}

\author{Wolfgang Belzig}
\affiliation{Fachbereich Physik, Universit{\"a}t Konstanz, D-78457 Konstanz, Germany}

\author{Juan Carlos Cuevas}
\affiliation{Departamento de F\'{i}sica Te\'{o}rica de la Materia Condensada, 
Universidad Aut\'{o}noma de Madrid, 28049 Madrid, Spain}
\affiliation{Condensed Matter Physics Center (IFIMAC), Universidad Aut\'{o}noma de Madrid, 28049 Madrid, Spain}

\date{\today}

\begin{abstract}
Multiterminal superconducting junctions have revitalized the investigation of the Josephson effect. One of the most
interesting aspects of these hybrid systems is the occurrence of multi-Cooper pair tunneling processes that have no
analog in two-terminal devices. Such correlated tunneling events are also intimately connected to the
Andreev bound states (ABSs) supported by these structures. Josephson junctions with four superconducting terminals 
have attracted special attention because they are predicted to support ABSs with nontrivial topological properties. 
Here, we present a theoretical study of \emph{sextets}, which are correlated tunneling processes involving three 
Cooper pairs and four different superconducting terminals. We investigate how sextets can be identified from 
the analysis of the current-phase relation, we show how sextets are connected to the hybridization of ABSs, and we 
discuss their existence in recent experiments on four-terminal devices realized in hybrid Al/InAs heterostructures.
\end{abstract}

\maketitle

\section{Introduction} \label{sec-intro}

The dc Josephson effect, i.e., the ability of a superconducting junction to sustain a nondissipative
current, is one of the most emblematic phenomena in the field of superconductivity \cite{Josephson1962,Barone1982}.
The modern view of the dc Josephson effect is that a superconducting phase difference across a Josephson junction
generates Andreev bound states (ABSs), which in turn mediate the Cooper pair transfer responsible of the 
supercurrent flow \cite{Beenakker1991,Furusaki1991}. The exact current-phase relation (CPR) of a superconducting
hybrid device depends on many details (materials, junction dimensions, etc.), and its connection with the ABS energy 
spectra has been widely investigated in two-terminal Josephson junctions (JJs), see 
Refs.~\cite{Pillet2010,Chang2013,Bretheau2013,Bretheau2013a,Janvier2015,Bretheau2017,vanWoerkom2017,Hays2018,Tosi2019,
Nichele2020,Hays2020,Hays2021,Pita-Vidal2023,Hinderling2023,Wesdorp2022} for recent examples.

The advent of multiterminal Josephson junctions (MTJJs) has provided a new impetus to the study of the Josephson 
effect and related issues \cite{Draelos2019,Graziano2020,Pankratova2020,Arnault2021,Graziano2022,Gupta2023}. 
The interest in these hybrid systems is manyfold. Thus, for instance, these MTJJs sustain
ABSs that depend on a number of superconducting phase differences, thus forming energy bands that may exhibit 
unique properties and eventually serve as a new type of qubits. In particular, it has been theoretically predicted 
that MTJJs are ideal platforms for band-structure engineering to realize topological systems 
\cite{Yokoyama2015,Riwar2016,Eriksson2017,Meyer2017,Xie2017,Xie2019,Repin2019,PeraltaGavensky2019,Houzet2019,Klees2020,
Weisbrich2021,Xie2022,Barakov2023,Teshler2023}. In fact, the first experimental steps in this direction have already
been reported making use of tunneling spectroscopy \cite{Coraiola2023,Antonelli2025}. Another topic of interest in 
the context of MTJJs is the use of these hybrid structures to investigate so-called Josephson diodes 
\cite{Coraiola2024}. Additionally, it has also been recently predicted that the ABSs in MTJJs could enable the 
study of engineered non-Hermitian topologies \cite{Ohnmacht2025}.

From the standpoint of quantum transport, MTJJs are also very interesting because they enable the exploration of 
complex tunneling processes that have no two-terminal analogs. A good example is the concept of a \emph{quartet}, which
consists in the simultaneous tunneling of two Cooper pairs originating in a given superconducting terminal and exiting the 
structure via two different electrodes \cite{Freyn2011,Jonckheere2013}. These tunneling events give rise to 
a very peculiar phase dependence in the CPR that reflects the correlated, multiterminal character of these processes 
\cite{Rech2014,Feinberg2015,Melo2022,Melin2023a}. Thus far, most of the experimental efforts to observe these quartet processes
have been based on probing the system response as a function of two bias currents or voltages 
\cite{Pfeffer2014,Cohen2018,Huang2022,Arnault2022}. A notable exception is Ref.~\cite{Ohnmacht2024} in which 
quartet signatures were found directly in the ABS tunneling spectra of a three-terminal device employing the so-called \emph{quartet tomography}, which consists of a Fourier analysis of either the CPR or the ABS band
structure. In that work, it was also shown that the occurrence of quartets is closely related to the hybridization 
of the ABSs. This fact illustrated that quartets are not only of fundamental interest, but also that their analysis 
serves to rigorously establish if a MTJJ exhibits genuine physics beyond that of two-terminal devices.

\begin{figure*}[t]
\includegraphics[width=0.75\textwidth,clip]{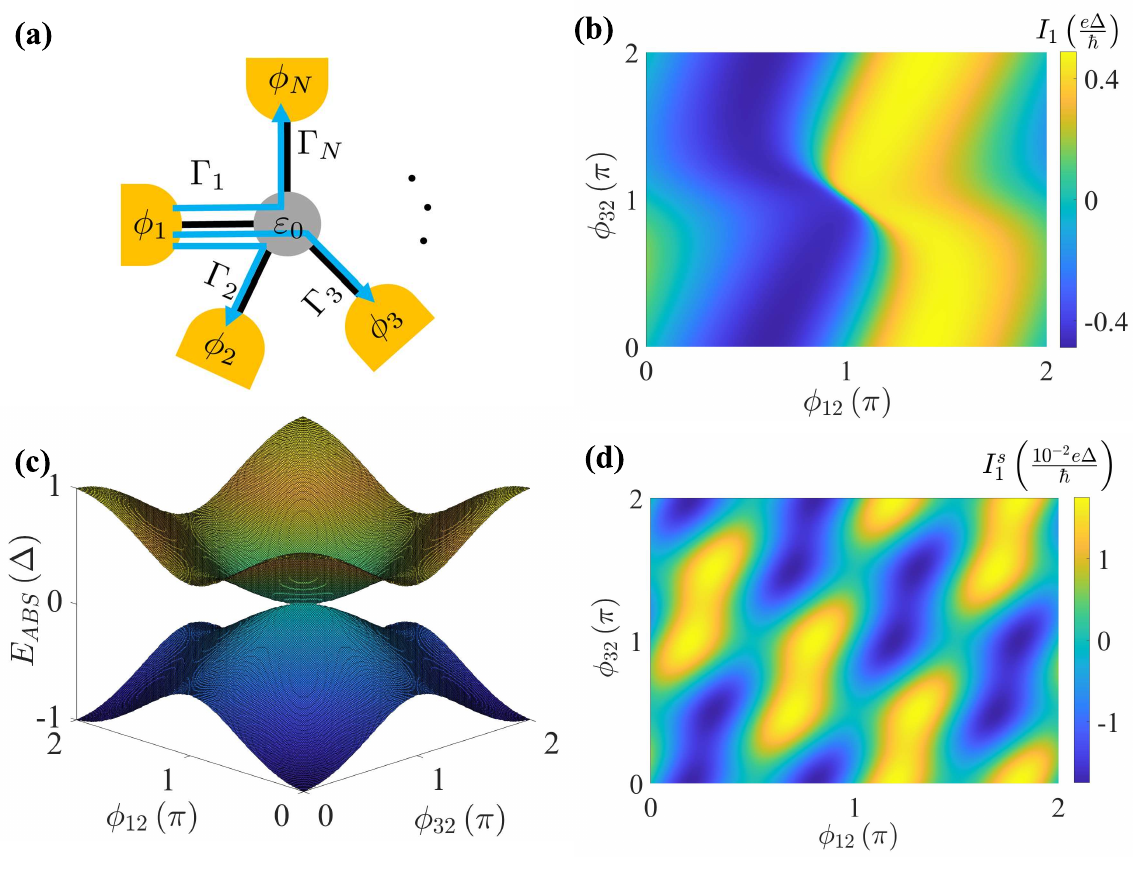}
\caption{(a) Schematics of the single-dot model. A single-level of energy $\epsilon_0$ is coupled to $N$ superconducting
terminals with phases $\phi_j$. The parameter $\Gamma_j$ describes the strength of the coupling between the dot and lead
$j$. The blue lines describe a sextet process in which 3 Cooper pairs tunnel simultaneously from terminal 1 and each
them leaves the hybrid structure via a different electrode. (b) Example of the current-phase relation 
$I_1(\phi_{12},\phi_{32},\phi_{42})$ for the model in (a) with four identical leads with energy gap $\Delta$. 
The different parameters are: $\Gamma_{j} = 5\Delta$ for $j\in\{1,2,3,4\}$, $\epsilon_0 = 0$, $\phi_4 = 0$, and 
$k_\text{B}T = 0.01\Delta$. (c) The corresponding phase dependence of the ABSs for the example in (b). (d) The total 
contribution of the four sextets to the current $I_1$ in the example of panel (b).}
\label{fig-1Dot}
\end{figure*}

The goal of this work is to extend the theoretical analysis of correlated tunneling processes to the case of
four-terminal devices. These MTJJs have become the holy grail of this field because, as mentioned above, it has 
been predicted that they can be used to realize topologically nontrivial phases in the three-dimensional ABS band
structure, in particular with the appearance of Weyl nodes \cite{Riwar2016}. In this context, we focus here on the
analysis of \emph{sextets}, which are processes that involve the simultaneous tunneling of three Cooper pairs that
involve up to four different superconducting terminals. To be precise, we focus here on short junctions, with
dimensions smaller than the superconducting coherence length. We make use of relatively simple models
to illustrate the nature of these tunneling events, to discuss under which conditions they can be observed in
the CPR, and we show how they are related to the hybridization of the ABSs in these MTJJs. Finally, we also discuss 
the possible existence of sextets in very recent experiments performed in four-terminal JJs realized in hybrid Al/InAs 
heterostructures \cite{Antonelli2025}. 

The rest of this manuscript is organized as follows. In Sec.~\ref{sec-1dot} we analyze a model of a 
four-terminal Josephson junction in which the normal region is made of a single quantum level. This simple
model allows us to illustrate the character of the sextets, as well as to discuss the conditions for their 
observation in the CPR. Section~\ref{sec-2dot-model} is devoted to the analysis of a double-dot model with the goal to 
elucidate the connection between the occurrence of sextets and the hybridization of the ABSs. In Sec.~\ref{sec-3dot}
we investigate the occurrence of sextets in a model that was recently employed to describe the ABS spectra in
four-terminal JJs realized in a hybrid Al/InAs heterostructure \cite{Antonelli2025}. We summarize the main
conclusions of this work in Sec.~\ref{sec-conclusions}. We also include two appendices to describe the 
technical details concerning the calculation of the supercurrent in the different models presented in this work,
see Appendix~\ref{appendix-GF}, and to provide additional results obtained with an insightful approximation,
see Appendix~\ref{appendix-Heff}.

\section{Sextet tomography: single-dot model} \label{sec-1dot}

Let us start our discussion by defining what we mean by a sextet and how it can be identified in the CPR of a MTJJ. 
For this purpose, it is didactic to analyze the simple model shown in Fig.~\ref{fig-1Dot}(a). In this model, 
a single-level, noninteracting quantum dot is coupled to $N$ superconducting leads. The energy of the spin-degenerate 
level is denoted by $\epsilon_0$ and the electrodes are $s$-wave superconductors with energy gaps $\Delta_j$ and 
superconducting phases $\phi_j$ ($j=1,\dots,N$). The coupling between the normal region and the lead $j$ is 
described by the tunneling rate $\Gamma_j$ (with dimensions of energy). Our objective is to calculate the 
nondissipative, zero-bias current flowing through the different terminals as a function of the superconducting 
phases $\phi_j$. To this end, we make use of Green's function techniques and our starting point are 
the (dimensionless) retarded and advanced Green's functions of the leads, which in a $2 \times 2$ Nambu representation 
read $\hat{g}^{\rm r/a}_j(E) = g^{\rm r/a}_j \hat{\tau}_0 + f^{\rm r/a}_j e^{\imath \phi_j \hat{\tau}_3}\hat{\tau}_1$, 
with $g^{\rm r/a}_j = -(E\pm \imath \eta)/\sqrt{\Delta_j^2-(E\pm \imath \eta)^2}$ and $f^{\rm r/a}_j =-\Delta_j /
\sqrt{\Delta_j^2-(E\pm \imath \eta)^2}$. Here, $E$ is the energy, $\eta = 0^+$, and $\tau_{0,1,3}$ are Pauli matrices. 
As shown in Appendix~\ref{appendix-GF}, one can write the current flowing through lead $i$ as $I_i = \sum_{j \neq i} 
I_{ij}$, where $I_{ij}$ is given by
\begin{equation} \label{eq_Iij}
    I_{ij} = \frac{8e}{h} \Gamma_i \Gamma_j \sin(\phi_{ji}) \int^{\infty}_{-\infty} \Im \left\{ 
    \frac{f^{\rm a}_i(E) f^{\rm a}_j(E)}{D(E,\boldsymbol{\phi})} \right\} n_{\rm F}(E) \, dE , 
\end{equation}
where $\phi_{ji} = \phi_j - \phi_i$, $n_{\rm F}(E)$ is the Fermi function, $\boldsymbol{\phi} = 
(\phi_1, \dots, \phi_N)$, and $D(E,\boldsymbol{\phi})$ is given by
\begin{eqnarray} \label{eq-D}
D(E,\boldsymbol{\phi}) & = & \left[ E - \epsilon_0 - \sum_k \Gamma_k g^{\rm a}_k \right]  
\left[ E + \epsilon_0 - \sum_k \Gamma_k g^{\rm a}_k \right] \nonumber \\ 
& & - \left[ \sum_k \Gamma_k f^{\rm a}_k e^{\imath \phi_k} \right]  
\left[ \sum_k \Gamma_k f^{\rm a}_k e^{-\imath \phi_k} \right] .
\end{eqnarray}
The ABS energies as a function of the different superconducting phases are obtained from the condition 
$D(E,\boldsymbol{\phi}) = 0$.

In this work, we are interested in four-terminal devices and for concreteness, we shall focus on the analysis of $I_1$,
the supercurrent flowing through terminal 1. This current depends on three phase differences, chosen $\phi_{12}$,  
$\phi_{32}$, and $\phi_{42}$, and can be written as the following Fourier series
\begin{equation} \label{eq_I1_1D}
    I_1(\phi_{12},\phi_{32},\phi_{42}) = 
    \sum_{l,n,m} c_{lnm} \sin (l \phi_{12} + n \phi_{32} + m \phi_{42}) .
\end{equation}

In a four-terminal Josephson junction, a sextet is a correlated tunneling process in which 3 Cooper pairs tunnel into
the normal region from a given terminal and each of them exits the structure through a different terminal.
This multi-Cooper pair tunneling process gives a contribution to the current that depends on a combination of 
phases of the type $\phi^{\rm s}_l = \phi_i + \phi_j + \phi_k - 3 \phi_l$. Here, $l$ corresponds to the
terminal from which the 3 Cooper pairs originate and all indices are different. In a four-terminal device, there are 
four sextets, each one originating from a different terminal. The contribution of sextet $l$ to $I_1$ is 
given by $S_l \sin(\phi^{\rm s}_l)$, where $S_1 = c_{-311}$, $S_2 = c_{111}$, $S_3 = c_{1-31}$, and 
$S_4 = c_{11-3}$. The first contribution, $S_1$, corresponds to the sextet in which three Cooper pairs 
enter the normal region from terminal 1 and exit separately via the other 3 terminals, see Fig.~\ref{fig-1Dot}(a). 
The other three sextets have similar interpretations. As shown in Appendix~\ref{appendix-GF}, one can do a perturbative
analysis of Eq.~\eqref{eq_Iij} to show that the Fourier coefficients determining the sextets are given to leading order 
in the $\Gamma$ parameters by: $c_{-311} = 3A_1, c_{111} = -A_{2}, c_{1-31} = -A_3, c_{11-3} = -A_4$, where 
\begin{eqnarray} \label{eq_Sl}
    A_l & = & \frac{4e}{h} \Gamma^3_l \Gamma_i \Gamma_j \Gamma_k \int^{\infty}_{-\infty}dE \, n_{\rm F}(E) \times \\
    & & \Im \left\{ \frac{ [f^{\rm a}_l(E)]^3 f^{\rm a}_i(E) f^{\rm a}_j(E) f^{\rm a}_k(E)}{(E^2 - \epsilon^2_0)^3} 
    \right\} . \nonumber
\end{eqnarray}
This equation supports the interpretation that a sextet of the type $\phi^{\rm s}_1 = \phi_2 + \phi_3 + \phi_4 -
3 \phi_1$ involves the injection of three Cooper pairs from terminal 1 that are transferred separately to leads 2, 3 and 4.
It is important to stress that, apart from these sextet contributions, the supercurrent $I_1$ has also contributions 
from all the typical two- and three-terminal processes. In any case, Eq.~\eqref{eq_I1_1D} provides a direct way 
to extract the sextet contributions, which simply consists in performing a Fourier analysis of the CPR \cite{Ohnmacht2024}. 
On the other hand, the form of Eq.~\ref{eq_I1_1D} is completely general for a four-terminal device, i.e., it is 
independent of the structure of the normal region, which only affects the individual coefficients $c_{lnm}$ in
Eq.~\ref{eq_I1_1D}.

\begin{figure}[b]
\includegraphics[width=\columnwidth,clip]{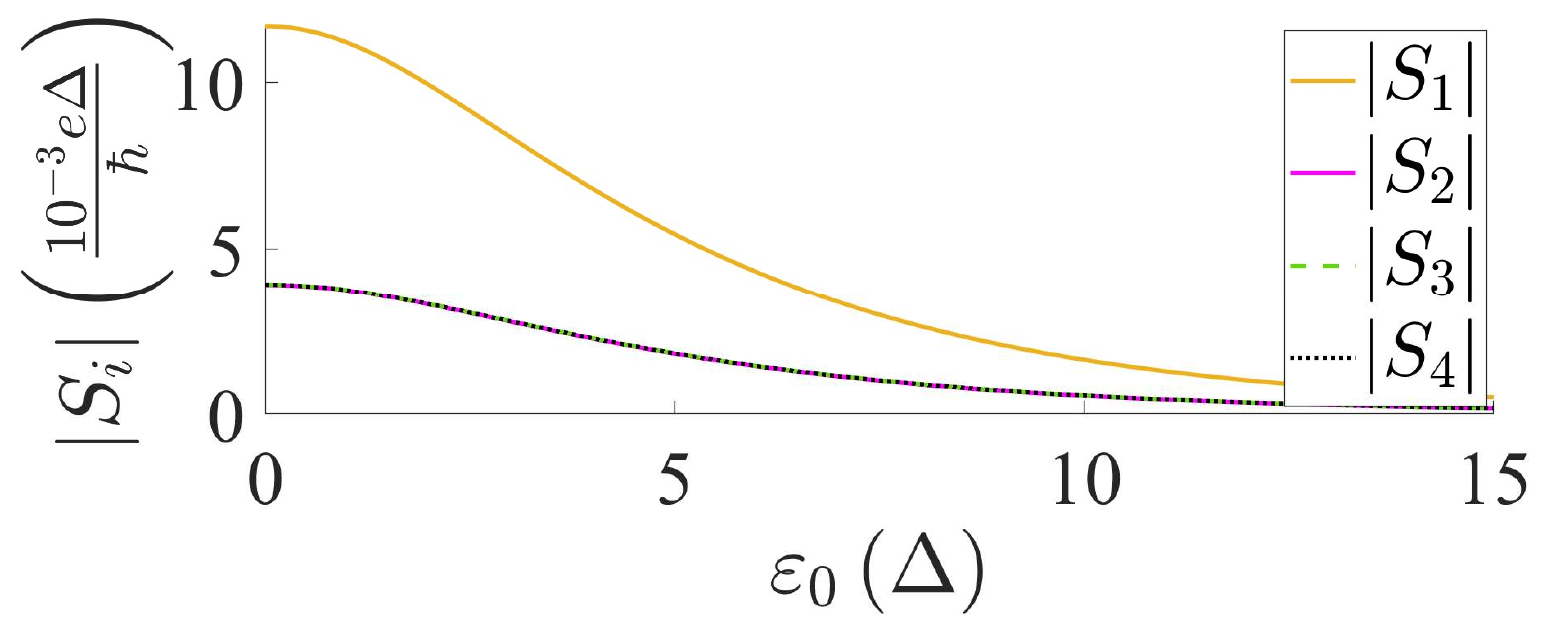}
\caption{Evolution of the magnitude of the sextet coefficients as a function of the level position in single-dot
model of Fig.~\ref{fig-1Dot} with four terminals. The parameters other than the level position are the same as in
Fig.~\ref{fig-1Dot}(b). Notice that there is a three-fold degeneracy of these coefficients.}
\label{fig-1DotS}
\end{figure}
\begin{figure*}[t!]
\includegraphics[width=0.75\textwidth,clip]{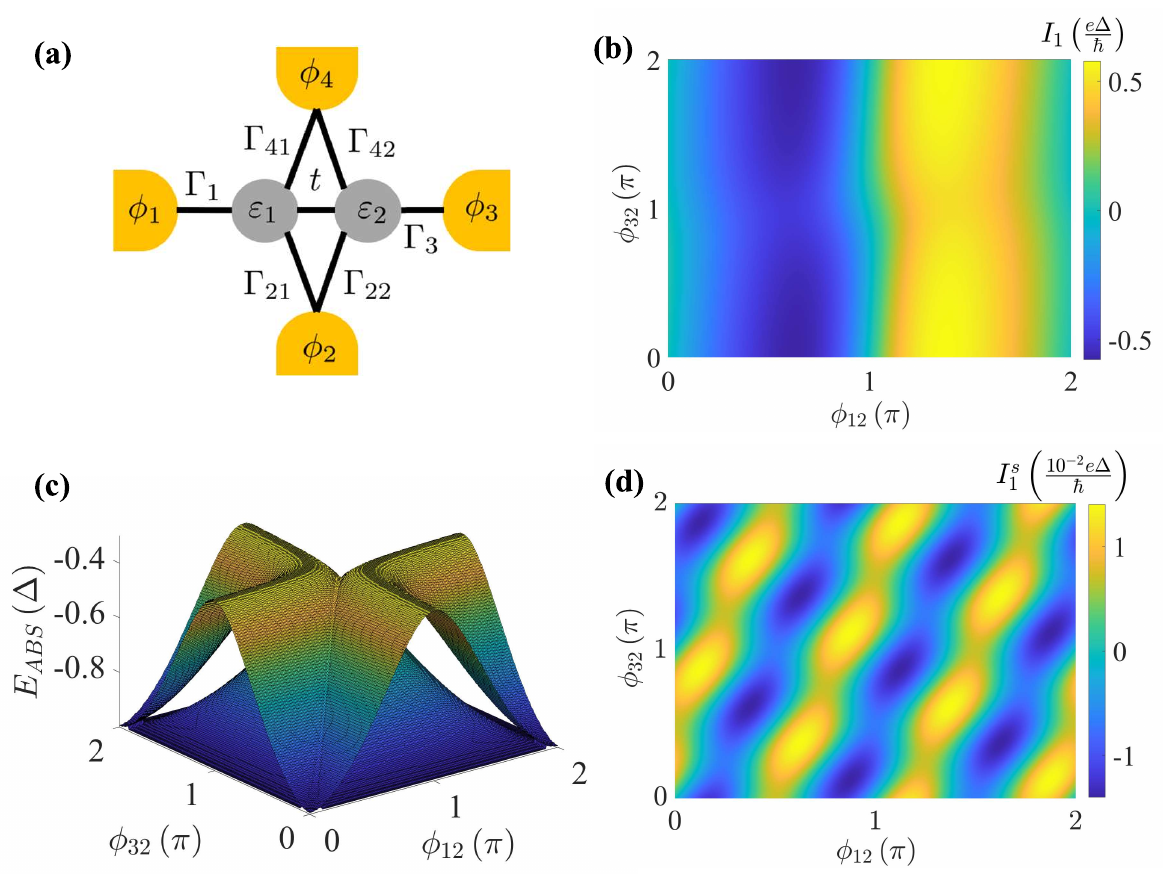}
\caption{(a) Schematics of the two-dot model. Two single-level quantum dots of energies $\epsilon_1$ and $\epsilon_2$ 
are coupled to four superconducting terminals with phases $\phi_j$. The parameters $\Gamma$ describe the strength of 
the coupling between the levels and the leads, as shown in this schematics. In particular, leads 2 and 4 are coupled 
to the two dots. The parameter $t$ describes the interdot coupling and controls the degree of hybridization of the ABSs.
(b) Example of the current-phase relation $I_1(\phi_{12},\phi_{32},\phi_{42})$ for the model in (a) with four 
identical leads with energy gap $\Delta$. The different parameters are: $\Gamma_{j} = 5\Delta$ for 
$j \in\{1,21,22,3,41,42\}$, $\epsilon_1 = \epsilon_2 = 0$, $t=5\Delta$, $\phi_4 = 0$, and $k_\text{B}T = 0.01\Delta$. 
(c) The corresponding phase dependence of the ABSs for the example in (b). We only show here the two states below the 
Fermi energy (there is electron-hole symmetry). (d) The total contribution of the four sextets to the current $I_1$ 
in the example of panel (b).}
\label{fig-2Dot}
\end{figure*}

We illustrate the results for this single-dot model in Fig.~\ref{fig-1Dot}(b) where we show the CPR $I_{\rm 1}$
for the parameters specified in the figure caption. On the other hand, and to illustrate the role of the sextets, we 
display in Fig.~\ref{fig-1Dot}(d) the supercurrent exclusively due to the four sextets, which was simply determined
by selecting the corresponding Fourier coefficients in Eq.~\eqref{eq_I1_1D}. Notice that these correlated tunneling 
processes give a non-negligible contribution to the total current. As mentioned above, it is important to emphasize 
that the sextet contributions are accompanied by many others, which include single- and multi-Cooper pair tunneling 
between pairs of terminals and quartets involving three terminals. This is the reason why these multi-Cooper pair 
tunneling processes are not easy to identify solely by exploring the critical current of these junctions. 

It is important to recall that in these coherent superconducting structures, the supercurrent is mainly carried 
by the ABSs generated by the superconducting phase difference. As stated above, the energies of those bound states 
can be obtained from the condition $D(E,\boldsymbol{\phi}) = 0$. In Fig.~\ref{fig-1Dot}(c) we show the ABSs
corresponding to the four-terminal example of panel (b). Notice that ABSs form a pair of energy bands with electron-hole
symmetry. In these MTJJ, the zero-temperature current $I_1$ is obtained from the energies $E^{(k)}_{\rm ABS}$ of the 
occupied ABSs as follows
\begin{equation} \label{eq-ABS-super}
	I_1(\boldsymbol{\phi}) = \frac{2e}{\hbar} \sum_k \frac{\partial 
    E^{(k)}_{\rm ABS}(\boldsymbol{\phi})}{\partial \phi_1} ,
\end{equation}  
where the factor 2 is due to spin degeneracy. We have verified that, in all the examples shown in this work, we can 
reconstruct the CPR using Eq.~\eqref{eq-ABS-super}. This demonstrates that the supercurrent is carried by the ABSs with 
no contributions from the continuum. As discussed in Ref.~\cite{Ohnmacht2024}, this connection between ABSs and supercurrent
suggests another protocol to identify sextets, which consists in the Fourier analysis of the ABS spectrum. That is a viable
strategy in those experiments in which the ABS spectrum is measured, e.g., via tunneling spectroscopy \cite{Coraiola2023}.
This was explicitly demonstrated in Ref.~\cite{Ohnmacht2024}. In this work, we shall not pursue this strategy and focus 
on the sextet signatures directly in the CPR.

Once we have established that sextets may exist in coherently coupled multi-terminal systems, the next natural issue
is to find out in which range of parameters these processes should be observable. Obviously, from the nature of these
multi-Cooper pair processes, one expects these tunneling events to give a sizable contribution to the current when these
structures are highly transparent. To illustrate this idea, we show in Fig.~\ref{fig-1DotS} the evolution of the sextet
Fourier coefficients with the level position $\epsilon_0$. By moving the level position away from the Fermi energy
($\epsilon_0=0$), the magnitude of these coefficients diminishes monotonically towards the tunnel regime, as expected. 
In this evolution, the ABS spectrum exhibit zero energy states when the level position lies at the Fermi energy, see 
Fig.~\ref{fig-1Dot}(d), and they move progressively towards the gap edges when the quantum dot level moves outside the 
gap (not shown here). The results shown in Fig.~\ref{fig-1DotS} also illustrate the fact that in the symmetric case in
which all dot-electrode couplings are equal, the sextet coefficients $S_2$, $S_3$, and $S_4$ are equal, as suggested by 
the analytical results of Eq.~\ref{eq_Sl}. Notice also that the quartet associated to the coefficient $S_1$ has a
magnitude that is 3 times larger that of the other three sextets, again as expected from the analytical results in 
this case.

\section{Sextets and ABS hybridization: two-dot model} \label{sec-2dot-model}

In most experimental situations there are several ABSs that significantly contribute to the supercurrent. Those ABSs,
in turn, may hybridize. This occurs when several terminals are coherently coupled. Since this is the situation in 
which sextets are expected to appear, one may wonder about the connection between ABS hybridization and sextets. To 
address this issue, we investigate now the simplest model that includes ABS hybridization, namely a two-dot system
coupled to four superconducting leads, see Fig.~\ref{fig-2Dot}(a). The key parameters of this model are described in this
figure and again, we neglect interaction effects in the quantum dots. The most important parameter for the following
discussion is the hybridization parameter or interdot coupling $t$, which in this four-terminal configuration controls 
the degree of hybridization between the two three-terminal ABSs states that may appear in this system (states involving 
terminals 1, 2, and 4 and 3, 2, and 4). The calculation of the supercurrent can be done in this case following the 
general recipe described in Appendix~\ref{appendix-GF}. Again, one can show that the generic CPR of Eq.~\ref{eq_I1_1D} 
applies to any four-terminal device and we have again four sextets, which can be identified in exactly the same way 
from the knowledge of the CPR.

In Fig.~\ref{fig-2Dot}(b), we show a particular example of the CPR in a high-transparent situation (see figure caption for
the list of parameters). We also show in panel (d) the corresponding contribution of the four sextets. Again, this
contribution is clearly sizable, providing a clear demonstration of the existence of this type of process. The 
corresponding ABSs for this example are shown in Fig.~\ref{fig-2Dot}(c), where only the two states below the Fermi 
energy are displayed (recall that there is always electron-hole symmetry in our systems).

\begin{figure}[t]
\includegraphics[width=\columnwidth,clip]{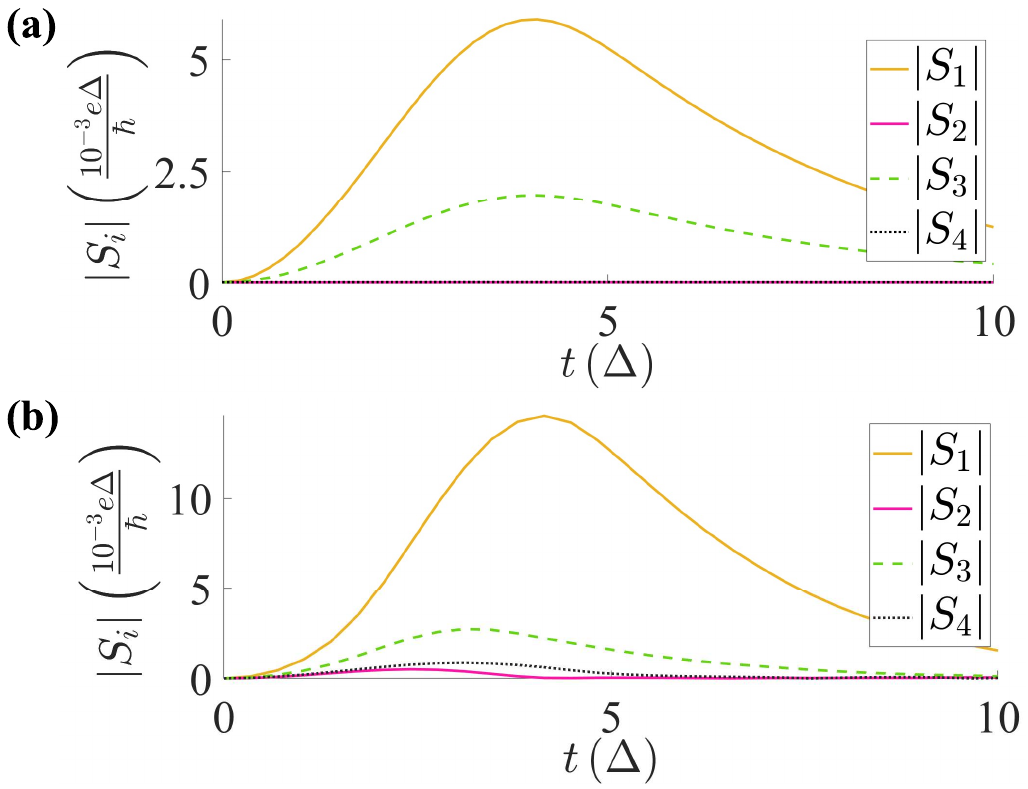}
\caption{(a) Evolution of the magnitude of the sextet coefficients as a function of the interdot coupling $t$ in the 
two-dot model of Fig.~\ref{fig-2Dot}(a). The parameters other than $t$ are the same as in Fig.~\ref{fig-2Dot}(b). 
(b) The same as in (a), but for asymmetric couplings: $\Gamma_1 = 5 \Delta, \Gamma_{21} = 2 \Delta, 
\Gamma_{22} = 3 \Delta, \Gamma_3 = 4 \Delta, \Gamma_{41} = 2.5 \Delta$ and $\Gamma_{42} = 4 \Delta$.}
\label{fig-2DotS}
\end{figure}

To address the issue on the connection between sextets and ABS hybridization, we explore the evolution of the 
magnitude of the sextets with the interdot hopping, $t$. This evolution is shown in Fig.~\ref{fig-2DotS}(a) for the
same parameters (other than $t$) as in Fig.~\ref{fig-2Dot}(b). There are several salient features. First, for $t=0$, 
i.e., when there is no hybrization between the ABSs, the sextet contributions exactly vanish, as expected. This
illustrates that the hybridization of the bound states requires the occurrence of sextets, or in other words,
the sextets are the origin of the ABS hybridization. Notice that the opposite is not true, i.e., the occurrence
of sextets does not necessarily imply ABS hybridization, as is obvious in the single-dot model of the previous section.
Second, for very large values of $t$ (larger than the couplings of the dots with the electrodes), the sextet amplitudes 
also tend to vanish. This is a simple consequence of the fact that in this limit a low transparent regime is achieved 
in which multi-Cooper pair tunneling events become very unlikely. Third, the nonvanishing coefficients reach their 
maximum around $t \sim \Gamma = 5 \Delta$. One can show that this situation corresponds to the highest transparent 
situation in which zero-energy ABSs become possible, due to a reflectionless scattering mode present in the normal state scattering matrix of the system \cite{Ohnmacht2025a}.

Finally, it is also worth noting that the sextets related to terminals 2 and 4 ($S_2$ and $S_4$) are 
exactly zero for the very symmetric configuration of Fig.~\ref{fig-2DotS}(a). We attribute this cancellation to 
destructive interference between the different paths that contribute to those two processes. This cancellation can be
explicitly shown in a perturbative expansion, where one finds that the contribution is only exactly zero, when the 
system if fully symmetric and the dot energies are exactly zero. To back this up, we have studied the evolution of 
the sextets in a situation in which the two couplings to terminals 2 and 4 are not identical, and the results are 
illustrated in Fig.~\ref{fig-2DotS}(b). Notice that now all sextets give a finite contribution (for $t \neq 0$), 
which supports our interpretation above.

\begin{figure*}[t]
\includegraphics[width=0.75\textwidth,clip]{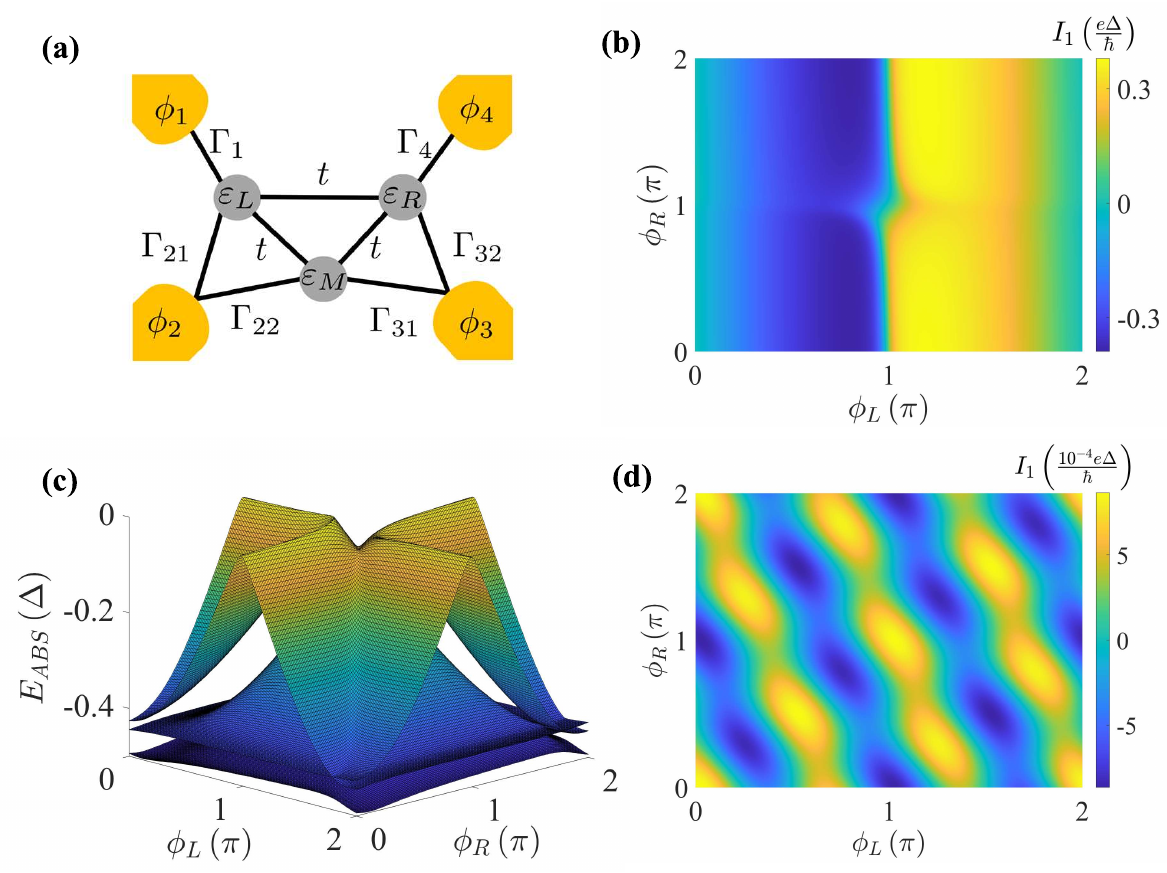}
\caption{(a) Schematics of the three-dot model taken from Ref.~\cite{Antonelli2025}. Three single-level quantum dots 
of energies $\epsilon_L$, $\epsilon_M$, and $\epsilon_R$ are coupled to four superconducting terminals with phases
$\phi_j$. We define $\phi_L = \phi_2-\phi_1$, $\phi_M = \phi_3-\phi_2$, and $\phi_R =\phi_4-\phi_3$. The parameters
$\Gamma$ describe the strength of the coupling between the levels and the leads, as shown in this schematics. The 
parameter $t$ describes the interdot coupling and controls the degree of hybridization of the ABSs. We assume that all
interdot couplings are equal. (b) Example of the current-phase relation $I_1$ for the model in (a) with four identical
leads with energy gap $\Delta$. The different parameters are: $\Gamma_{j} = 0.37\Delta$ for $j \in\{1,21,22,3,41,42\}$,
$\epsilon_L = \epsilon_M = \epsilon_R = 0.005\Delta$, $t = 0.12\Delta$, $\phi_M = 0.21\pi$, and 
$k_\text{B}T = 0.01\Delta$. (c) The corresponding phase dependence of the ABSs for the example in (b). We only show 
here the three states below the Fermi energy (there is electron-hole symmetry). (d) The total contribution of the four
sextets to the current $I_1$ in the example of panel (b).}
\label{fig-3Dot}
\end{figure*}

\section{Sextets in a realistic setup} \label{sec-3dot}

The previous sections were based on simple models that allowed us to illustrate the nature of sextets and their
connection to the ABS hybridization. The last issue that we would like to address is the possibility to have 
sextets in realistic setups. For this purpose, we investigate now the model described in Fig.~\ref{fig-3Dot}(a)
which we recently employed to describe the results of the tunneling spectroscopy in a four-terminal device 
fabricated in a hybrid Al/InAs heterostructure \cite{Antonelli2025}. In that work, the first tunneling spectroscopy 
of ABSs in a four-terminal device was reported in which three different superconducting phases were independently 
controlled. Furthermore, signatures of an Andreev tri-molecule were found where ABSs depend on all three superconducting 
phase differences. On the other hand, the model of Fig.~\ref{fig-3Dot}(a) containing three dots was shown to
explain the key experimental observations concerning the phase dependence of the three low-energy ABSs exhibited by
this four-terminal Josephson junction. Moreover, this model was used to argue that under the experimental conditions 
a phase-controlled topological phase transition could take place. The question arises: does this model predict 
the occurrence of sextets in this MTJJ?

We show in Fig.~\ref{fig-3Dot}(b) an example of the CPR for the values used in Ref.~\cite{Antonelli2025} to
describe the experimental results. We have slightly modified our notation to make it consistent with that of 
Ref.~\cite{Antonelli2025} and we now express the CPR in terms of the following three phase differences: 
$\phi_L = \phi_2-\phi_1$, $\phi_M = \phi_3-\phi_2$, and $\phi_R =\phi_4-\phi_3$. Notice that this CPR only differs 
from a CPR of two independent ABSs in the region around $(\pi,\pi)$, in which the states are expected 
to hybridize the most. The corresponding ABSs computed with this model and for the same set of parameters are shown 
in Fig.~\ref{fig-3Dot}(c). Notice in particular that the highest-energy state only shows signs of hybridization in 
the region $(\pi,\pi)$ mentioned above, which explains the behavior of the CPR. More importantly, Fig.~\ref{fig-3Dot}(d) 
displays the corresponding contribution of the sextets to the CPR. These results demonstrate that such processes are present in this multiterminal junction, which further confirms the formation of an Andreev tri-molecule demonstrated 
in Ref.~\cite{Antonelli2025}.

It is interesting to mention that the low-energy physics of this model can be studied with the help of an effective
Hamiltonian, which can be derived from the full Green's function approach by simply focusing on low energies, see
Appendix~\ref{appendix-Heff} for details. This low-energy approximation is convenient, for instance, to analyze the
topological nature of the ABSs, as shown in Ref.~\cite{Antonelli2025}. Here, we have used this semi-analytical 
approach to confirm the main conclusions above for the CPR, the contribution of the sextets and the phase-dependence 
of the ABSs, see details in Appendix~\ref{appendix-Heff}.

\begin{figure}[t]
\includegraphics[width=\columnwidth,clip]{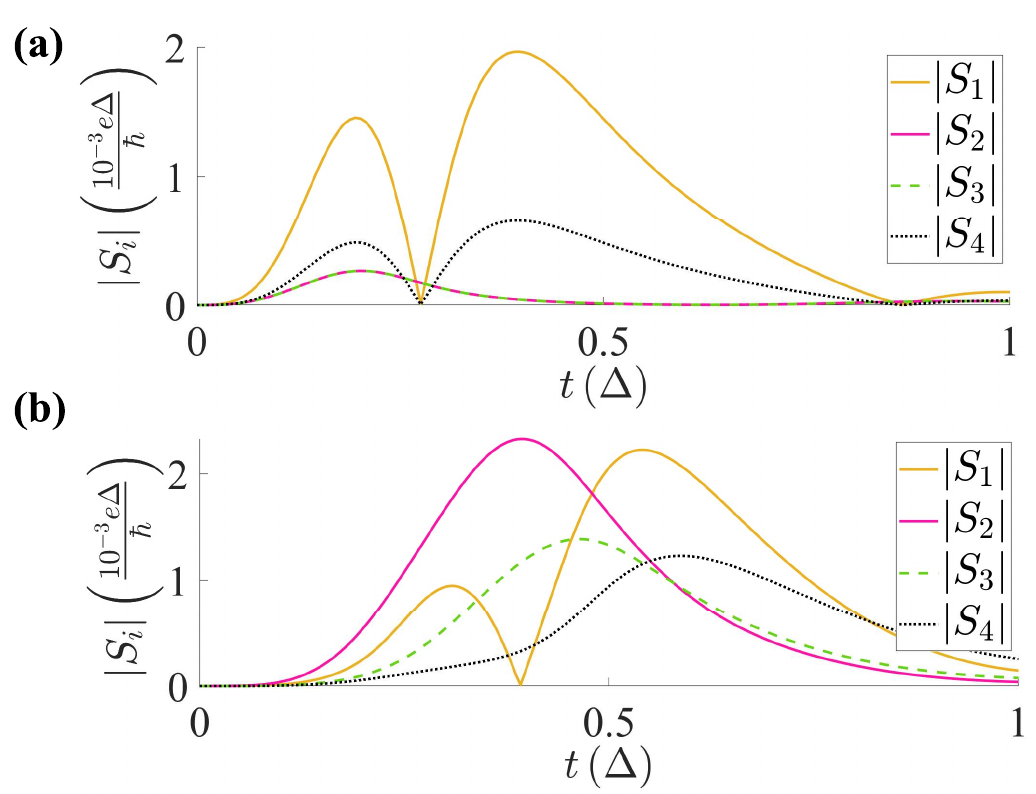}
\caption{(a) Evolution of the magnitude of the sextet coefficients as a function of the interdot coupling $t$ in the 
three-dot model of Fig.~\ref{fig-3Dot}(a). The parameters other than $t$ are the same as in Fig.~\ref{fig-3Dot}(b). 
(b) The same as in (a), but for asymmetric couplings:  $ \Gamma_1= 0.37 \Delta, \Gamma_{21} = 0.5 \Delta, 
\Gamma_{22} = 0.25 \Delta, \Gamma_{31} = 0.75 \Delta, \Gamma_{32} = 0.45 \Delta$ and $\Gamma_4 = 0.8 \Delta$.}
\label{fig-3DotS}
\end{figure}

For completeness, we have also studied for this model the evolution of the magnitude of the sextets as a function of 
the hybridization parameter. We show in Fig.~\ref{fig-3DotS}(a) this evolution for the set of parameters of the
example in Fig.~\ref{fig-3Dot}(b). As expected, the sextet contributions tend to vanish in the two limiting
cases: $t \to 0$ and $t \to \infty$. Notice that the evolution of these coefficients is not completely
monotonic, again suggesting the impact of interference effects. Notably, the contributions $S_{1/4}$ have a sign 
change in their amplitude and go to zero at a non-zero value of the hybridization. Interestingly, in the vicinity 
of this hybridization value, there is a topological phase boundary where Weyl nodes originate \cite{Ohnmacht2025a}, 
indicating that there might be a direct connection between sextet contributions and the underlying topology. It 
should also be noted that two of the coefficients are degenerate ($S_2$ and $S_3$). Again, this degeneracy can be 
lifted by asymmetrizing the couplings to those two leads, as we demonstrate in Fig.~\ref{fig-3DotS}(b).

\section{Conclusions} \label{sec-conclusions}

In summary, motivated by recent experiments, we have presented in this work a theoretical analysis of the occurrence 
of sextets in four-terminal Josephson junctions. The analysis of these multi-Cooper pair tunneling processes provides 
an unambiguous way, at least from the theory side, to determine whether a MTTJ exhibits genuine multiterminal physics.
Using simple models in combination with Green's function techniques, we have studied the conditions for the occurrence 
of these processes and how they can be identified in the analysis of the CPR. We have also shown that they are closely
related to the hybridization of ABSs in these heterostructures. Moreover, we have studied the possible occurrence of 
sextets in recent experiments realized in hybrid Al/InAs heterostructures \cite{Antonelli2025}. Our analysis supports 
the idea that such four-terminal devices indeed feature sextets.

Our study here can be extended in many different directions. For instance, following Ref.~\cite{Ohnmacht2024},
it is straightforward to analyze the impact of sextets in the ABS spectra of these four-terminal superconducting
devices. Another interesting topic is the possible influence of the spin-orbit interaction in these multi-Cooper
pair tunneling processes, an interaction that under certain circumstances can indeed affect the ABS spectra of
hybrid structures based on semiconductor two-dimensional electron gases \cite{Coraiola2023a}. Most interestingly, 
the connection between non-trivial topology and sextet processes is still not clear. Considering that non-trivial 
topology only becomes possible in true 4-terminal MTJJs, the existence of sextet processes is a necessary condition 
for the existence of topology. Furthermore, we find direct signatures that for certain sextet contributions, a sign 
change with a zero crossing appears which might be related to topological phase boundaries. Exploring this circumstance
further could provide a useful tool for experimentalists in investigating the possible connection of the topological 
properties of the ABS spectra and sextets. Last but not least, it would also be desirable to investigate the role of 
interactions in the occurrence of sextets and related processes.

\begin{acknowledgments}
We acknowledge fruitful discussions with Valentin Wilhelm.
D.C.O.\ and W.B.\ acknowledge support by the Deutsche Forschungsgemeinschaft (DFG; German Research Foundation) via SFB 1432 (Project No. 425217212). J.C.C.\ thanks the DFG and SFB 1432 for sponsoring his stay at the University of Konstanz as a Mercator Fellow.
\end{acknowledgments}

\appendix

\section{Green's function formalism: Supercurrent and ABSs} \label{appendix-GF}

In this appendix, we provide the technical details for the calculation of the supercurrent and the ABS spectrum
in the different models discussed in the main text. To give a unified description that can be applied to all these
models, we consider a generic MTJJ with $N$ superconducting leads coupled by a central normal region formed by
$M$ interconnected single-level quantum dots. Such a system can be described by the following Hamiltonian 
\begin{align}\label{f: HS1}
    \hat{H} = \sum_{j =1}^N \hat{H}_{\text{L},j} + \hat{H}_\text{C} + \hat{H}_\text{LC} .
\end{align}
Here, $\hat{H}_{\text{L},j}$ is the BCS Hamiltonian of lead $j$ given by
\begin{align} \label{f: HLead}
    \hat{H}_{\text{L},j} &= \sum_{\mathbf{k},\sigma} \epsilon_{\mathbf{k}j}\, \hat{c}^\dagger_{\mathbf{k}j\sigma} 
    \hat{c}_{\mathbf{k}j\sigma} \\ \nonumber
&+ \sum_{\mathbf{k}} \left( |\Delta_j| e^{i\phi_j}\, \hat{c}^\dagger_{\mathbf{k}j\uparrow} 
\hat{c}^\dagger_{-\mathbf{k}j\downarrow}
+ |\Delta_j| e^{-i\phi_j}\, \hat{c}_{-\mathbf{k}j\downarrow} \hat{c}_{\mathbf{k}j\uparrow} \right) ,
\end{align} 
where $\hat{c}^\dagger_{\mathbf{k}j\sigma}$ and $\hat{c}_{\mathbf{k}j\sigma}$ are the creation and annihilation 
operator, respectively, of an electron in state $\mathbf{k}$, in lead $j$ and with spin $\sigma$. Moreover,
$|\Delta_j|$ is the magnitude of the order parameter of lead $j$ and $\phi_j$ is the corresponding phase.
On the other hand, $\hat{H}_\text{C}$ is the Hamiltonian of the central region that adopts the form
\begin{equation} \label{eq-HC}
    \hat{H}_\text{C} = \sum_{i, \sigma} \varepsilon_i \hat{d}_{i \sigma}^\dagger \hat{d}_{i \sigma} + 
    \sum_{ij,\sigma} t_{ij} \hat{d}_{i\sigma}^\dag \hat{d}_{j\sigma} ,
\end{equation} 
where $\hat{d}_{i \sigma}^\dagger$ and $\hat{d}_{i \sigma}$ are the creation and annihilation operators,
respectively, of an electron in quantum dot $i$ and spin $\sigma$. By defining $\tilde{d}^\dag =
\left(\hat{d}_{1\uparrow}^\dag, \hat{d}_{1\downarrow},...,\hat{d}_{M\uparrow}^\dag, \hat{d}_{M\downarrow}\right)$, 
we can rewrite the Hamiltonian of the central region as $\hat{H}_\text{C} = \tilde{d}^\dagger \bar{H}_\text{C} 
\tilde{d}$, where $\bar{H}_\text{C}$ is given by 
\begin{align}
    \bar{H}_\text{C} = \begin{pmatrix}
    \epsilon_1 & 0 & t_{12} & 0 & \cdots & t_{1M} & 0 \\
    0 & -\epsilon_1 & 0 & -t_{12} & \cdots & 0 & -t_{1M} \\
    t_{21} & 0 & \epsilon_2 & 0 & \cdots & t_{2M} & 0 \\
    0 & -t_{21} & 0 & -\epsilon_2 & \cdots & 0 & -t_{2M} \\
    \vdots & \vdots & \vdots & \vdots & \ddots & \vdots & \vdots \\
     t_{M1} & 0 & \cdots& t_{MM-1} & 0  & \epsilon_M & 0 \\
    0 & - t_{M1} & \cdots & 0 & -t_{MM-1} & 0 & -\epsilon_M \\
    \end{pmatrix} .
\end{align}
Finally, $\hat{H}_\text{LC}$ describes the coupling between the central region and the superconducting leads
and reads
\begin{equation} \label{eq-HLC}
    \hat{H}_\text{LC} = \sum_{i j, \textbf{k}, \sigma} w_{i j}\left(\hat{d}_{i \sigma}^\dagger 
    \hat{c}_{\textbf{k}j\sigma}+ \hat{c}_{\textbf{k}j\sigma}^\dagger \hat{d}_{i\sigma} \right) ,
\end{equation} 
where $w_{ij}$ is the tunneling matrix element between lead $i$ and dot $j$, which we assume to be real
and independent of $\textbf{k}$. Related to these matrix elements, we define the tunneling rates 
$\Gamma_{ij} = \pi N_{0i} w_{ij}^2$, where $N_{0i}$ is the normal density of states at the Fermi energy of 
lead $i$. 

Our goal now is to compute the supercurrent flowing, for instance, through terminal 1. The corresponding 
current density operator is given by
\begin{equation} \label{eq-I1-op}
    \hat I_1(t)  = \frac{ie}{\hbar} \sum_{\textbf{k}, j, \sigma} w_{1 j} 
    \left( \hat{c}_{\textbf{k}1\sigma}^\dagger(t) \hat{d}_{j\sigma}(t) - 
    \hat{d}_{j \sigma}^\dagger(t) \hat{c}_{\textbf{k}1\sigma}(t) \right) .
\end{equation} 

Next, we use the definition of the lesser Green's functions within the Nambu representation:
\begin{equation}
    \hat G^{+-}_{ij}(t, t^{\prime}) = -i \sum_{\textbf{k}} \left\langle \hat{T}_C \left\{ 
    \tilde{c}_{\textbf{k} i}(t_+)\, \tilde{d}^\dagger_{j}(t^{\prime}_-) \right\} \right\rangle ,
\end{equation}
where $\hat{T}_C$ is the time-ordering operator on the Keldysh contour and we have defined the
spinors $\tilde{c}^{\dagger}_{\textbf{k}i} = (c^{\dagger}_{\textbf{k}i \uparrow}, c_{-\textbf{k}i \downarrow})$ 
and $\tilde{d}^{\dagger}_j = (d^{\dagger}_{j \uparrow}, d_{j \downarrow})$. In terms of these Green's 
functions, the expectation value $I_1 = \langle \hat{I}_1 \rangle$ can be written as
\begin{equation} \label{eq-I1}
    I_1 =  \frac{e}{\hbar} \sum_{j} \text{Tr} \left\{ \hat{\tau}_3 
    \left( \hat{V}_{j1} \hat{G}_{1j}^{+-}(t,t)- \hat{G}_{j1}^{+-}(t,t) \hat{V}_{1j}\right)\right\} ,
\end{equation} 
where $\text{Tr}$ denotes the trace in Nambu space, $\hat{\tau}_3$ is the third Pauli matrix in that space,
and we have defined
\begin{equation}
\hat{V}_{1j} = \hat{V}_{j1}= 
\begin{pmatrix} w_{1j} & 0\\  0 &  -w_{1j} \end{pmatrix} .
\end{equation}

Since we are interested in the supercurrent, the system is in equilibrium and the Green's functions only depend on the 
difference of the time arguments. Thus, we can Fourier transform the Green's functions to energy space and rewrite 
the current as
\begin{equation}
I_1 = \frac{e}{h} \sum_j \int^{\infty}_{-\infty} dE \, \mbox{Tr} \left\{ \hat{\tau}_3 
\left[ \hat V_{j1} \hat G^{+-}_{1j}(E) - \hat G^{+-}_{j1}(E) \hat V_{1j} 
\right] \right\} .
\end{equation}
Moreover, we can make use of the equilibrium relation
\begin{equation} \label{eq-G+-}
\hat G^{+-}(E) = [ \hat G^{\rm a}(E) - \hat G^{\rm r}(E) ] n_{\rm F}(E) ,	
\end{equation}
where $n_{\rm F}(E)$ is the Fermi function, to rewrite the lesser Green's functions in terms of the retarded and
advanced ones.

\begin{figure*}[t]
\includegraphics[width=\textwidth,clip]{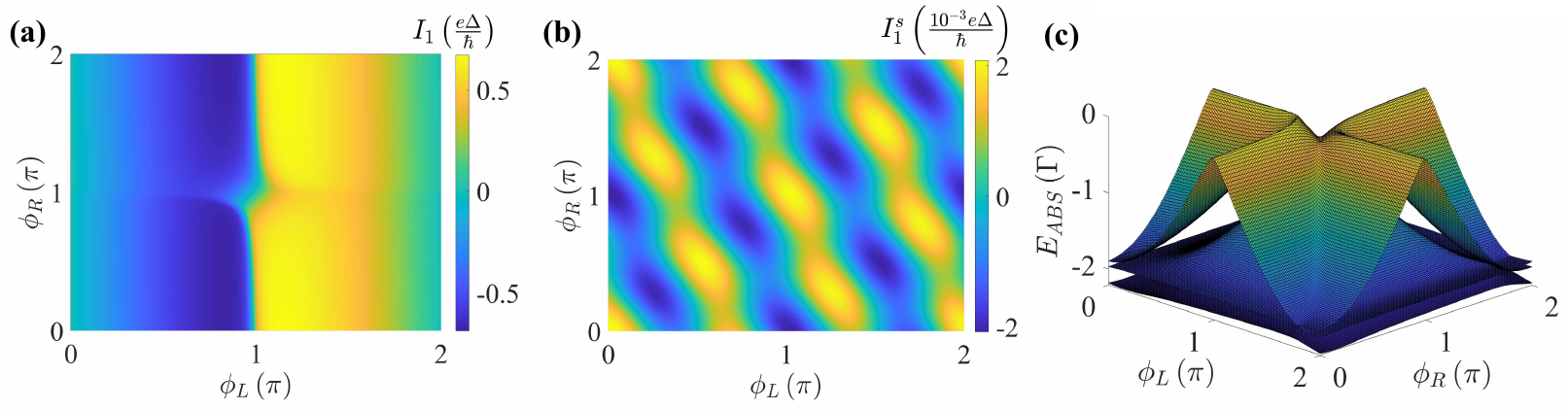}
\caption{(a) Current-phase relation $I_1$ for the model in Fig.~\ref{fig-3Dot}(a) using the effective Hamiltonian 
approximation. The different parameters are those used in Fig.~\ref{fig-3Dot}(b). 
(b) The total contribution of the four sextets to the current $I_1$ in the example of panel (b) using the
effective Hamiltonian approximation. (c) The phase dependence of the ABSs for the example in (a)
obtained diagonalizing the corresponding effective Hamiltoninan. We only show here the three 
states below the Fermi energy (there is electron-hole symmetry).}
\label{fig-3Dot-Heff}
\end{figure*}

To compute the different Green's functions we follow a perturbative approach in which we assume that the whole 
system is divided into different subsystems, namely the leads and the central normal region formed by the array 
of dots. We consider the couplings between leads and the central system as the perturbation. With this choice, 
the Green's functions can be calculated by solving the corresponding Dyson's equations. Thus, for instance, we can 
compute the retarded and advanced Green's functions that enter into the current expression using the following Dyson's equations
\begin{equation}
 \hat G^{\rm r,a}_{j1} = \sum_i \hat G^{\rm r,a}_{ji} \hat V_{i1} \hat g^{\rm r,a}_{1} \;\;\; \mbox{and} \;\;\;
 \hat G^{\rm r,a}_{1j} = \sum_i \hat g^{\rm r,a}_{1} \hat V_{1i} \hat G^{\rm r,a}_{ij} ,
\end{equation}
where $\hat g$ denotes the unperturbed Green's functions. Using these relations, together with Eq.~\eqref{eq-G+-}, 
we can rewrite the current as
\begin{equation} \label{eq-I-1D-int}
	I_1 = \frac{e}{h} \int^{\infty}_{-\infty} \mbox{Tr}^{\prime} \left\{ \bar{\tau}_3 \left( 
	\left[ \bar \Sigma^{\rm a}_1 , \bar G^{\rm a}_{\text{C}} \right] - 
    \left[ \bar \Sigma^{\rm r}_1 , \bar G^{\rm r}_{\text{C}} \right] \right) \right\}n_{\rm F}(E) \, dE ,
\end{equation}
where $(\, \bar{\,} \,)$ indicates that the corresponding quantity is a $2M \times 2M$ matrix in the 
combined Nambu and dot space and $\mbox{Tr}^{\prime}$ represents the trace in this combined space. Moreover, 
$\bar{\tau}_3 = \hat{\tau}_3 \otimes \hat{1}_M$, where $\hat{1}_M$ is the $M \times M$ unit matrix, and the different 
$2 \times 2$ blocks of the self-energy $\bar \Sigma^{\rm r,a}_{l}$, corresponding to lead $l$, are defined as
\begin{equation}
	[\bar \Sigma^{\rm r,a}_l]_{ij}(E) = \hat V_{il} \hat g^{\rm r,a}_{l}(E) \hat V_{lj} .
\end{equation}
Here, $\hat g^{\rm r,a}_{l}$ are the Green's function of lead $l$ given by 
\begin{eqnarray} \label{eq-gj}
    \hat{g}^{\rm r,a}_{l}(E) & \equiv & \begin{pmatrix} g^{\rm r,a}_l & f^{\rm r/a}_l e^{\imath \varphi_j} \\ 
    f^{\rm r,a}_l e^{-\imath \varphi_l} & g^{\rm r,a}_l  \end{pmatrix} \\
    & = & \frac{-1}{\sqrt{|\Delta_l|^2-(E\pm \imath \eta)^2}} \begin{pmatrix} E\pm \imath \eta & 
    |\Delta_l| e^{\imath \varphi_l} \\ 
    |\Delta_l| e^{-\imath \varphi_l} & E\pm \imath \eta \end{pmatrix} , \nonumber
\end{eqnarray}
where $E$ is the energy and $\eta = 0^+$.

Using $\hat G^{\rm r} = (\hat G^{\rm a} )^{\dagger}$, we can express Eq.~\eqref{eq-I-1D-int} as
\begin{equation} \label{eq-I1-gen}
	I_1 = \frac{2e}{h} \int^{\infty}_{-\infty} \Re \left( \mbox{Tr}^{\prime} \left\{ \bar{\tau}_3 
	\left[ \bar \Sigma^{\rm a}_1 , \bar G^{\rm a}_{\text{C}} \right] \right\} \right) n_{\rm F}(E) \, dE .
\end{equation}
The last step now is to compute $\hat G^{\rm a}_{\text{C}}$, which can be obtained from the following Dyson's equation
\begin{equation} \label{eq-Dyson-GCC}
	\hat G^{\rm a}_{\text{C}} = \left[ (E - i\eta) \bar{1} - \bar{H}_{\text{C}} - 
    \sum_{l} \bar \Sigma_l \right]^{-1} .
\end{equation}

This is the general recipe that can be applied to any of the models in the main text. In the case of the 
single-dot model of Sec.~\ref{sec-1dot}, it is straightforward to show that this recipe leads to Eqs.~\eqref{eq_Iij} 
and \eqref{eq-D} for the supercurrent through lead 1.

Let us now briefly discuss how to identify the sextet contributions in the CPR, focusing on the single-dot model
of Sec.~\ref{sec-1dot} with four terminals. The denominator that appears in the current expression of 
Eq.~\eqref{eq_Iij}, which is defined in Eq.~\eqref{eq-D}, can be written as $D = (E^2-\epsilon^2_0) [1 -x]$, where
\begin{eqnarray}
	x & = & \frac{1}{(E^2-\epsilon^2_0)} \left\{ \sum_j \Gamma^2_j + 2E \sum_j \Gamma_j g^{\rm a}_j + \right.  
	\nonumber \\ & & \left. 2 \sum_{\langle jk \rangle} \Gamma_j \Gamma_k [ f^{\rm a}_j f^{\rm a}_k \cos (\phi_{jk}) - 
	g^{\rm a}_j g^{\rm a}_k ] \right\} ,
\end{eqnarray}
Here, $\sum_{\langle jk \rangle}$ indicates a sum over all possible pair of leads with $j \neq k$. 
Thus, $1/D$ can be written
as
\begin{equation}
	\frac{1}{D} = \frac{1}{(E^2-\epsilon^2_0)} \frac{1}{1-x} = \frac{1}{(E^2-\epsilon^2_0)}
	\sum^{\infty}_{n=0} x^n .
\end{equation}
The terms in the current proportional to $x^2$ include the sextet contributions to the leading order (sixth order 
in the $\Gamma$'s). In particular, after straightforward manipulations, one can identify the following contributions 
\begin{eqnarray}\label{f: 4SLead1NDot_I1Short}
    I_1^{s} & = & 3A_1\text{sin}(\phi_2+\phi_3+\phi_4-3\phi_1) \\
    & & - A_2\text{sin}(\phi_1+\phi_3+\phi_4-3\phi_2) \nonumber\\
    & & - A_3\text{sin}(\phi_1+\phi_2+\phi_4-3\phi_3) \nonumber\\
    & & - A_4\text{sin}(\phi_1+\phi_2+\phi_3-3\phi_4) , \nonumber
\end{eqnarray} 
where the coefficients $A_i$ are defined in Eq.~\eqref{eq_Sl}.

\section{Effective Hamiltonian approximation} \label{appendix-Heff}

As mentioned in the main text, it is often convenient to derive an effective Hamiltonian to describe the low-energy
physics of all the models considered in this work. In particular, that is very useful for the analysis of the 
topological properties of the ABSs, see e.g.\ Ref.~\cite{Antonelli2025}. This effective Hamiltonian can be 
derived from Eq.~\eqref{eq-Dyson-GCC} by simply taking the limit $E \to 0$ as follows
\begin{equation}
    \bar{H}_{\text{eff}} = \bar{H}_{\text{C}} + \sum_j \bar{\Sigma}_j(E=0) .
\end{equation}
From this Hamiltonian, the ABS spectrum can be obtained by diagonalizing it, while the CPR can be computed from
the knowledge of the phase-dependence of the ABS energies by using Eq.~\eqref{eq-ABS-super}.

Obviously, this approximation cannot be expected to reproduce the physics of the ABSs for high energies 
close to the gap. In any case, it provides useful insight and allows us to test the basic properties of both the
CPR and the ABS spectra. As an example, we show in Fig.~\ref{fig-3Dot-Heff} the results for both the CPR and the ABSs 
for the three-dot model of Sec.~\ref{sec-3dot} and for the same parameters as in Fig.~\ref{fig-3Dot}. Notice that 
in this case the energy of the ABSs is normalized by the coupling $\Gamma$ since the gap $\Delta$ is considered
to be infinite in this approximation. In any case, one can see that this approximation nicely reproduces all the
salient features discussed in Sec.~\ref{sec-3dot}.

\bibliographystyle{apsrev4-2}
\bibliography{MyLibrary}

\begin{thebibliography}{56}%
\makeatletter
\providecommand \@ifxundefined [1]{%
 \@ifx{#1\undefined}
}%
\providecommand \@ifnum [1]{%
 \ifnum #1\expandafter \@firstoftwo
 \else \expandafter \@secondoftwo
 \fi
}%
\providecommand \@ifx [1]{%
 \ifx #1\expandafter \@firstoftwo
 \else \expandafter \@secondoftwo
 \fi
}%
\providecommand \natexlab [1]{#1}%
\providecommand \enquote  [1]{``#1''}%
\providecommand \bibnamefont  [1]{#1}%
\providecommand \bibfnamefont [1]{#1}%
\providecommand \citenamefont [1]{#1}%
\providecommand \href@noop [0]{\@secondoftwo}%
\providecommand \href [0]{\begingroup \@sanitize@url \@href}%
\providecommand \@href[1]{\@@startlink{#1}\@@href}%
\providecommand \@@href[1]{\endgroup#1\@@endlink}%
\providecommand \@sanitize@url [0]{\catcode `\\12\catcode `\$12\catcode `\&12\catcode `\#12\catcode `\^12\catcode `\_12\catcode `\%12\relax}%
\providecommand \@@startlink[1]{}%
\providecommand \@@endlink[0]{}%
\providecommand \url  [0]{\begingroup\@sanitize@url \@url }%
\providecommand \@url [1]{\endgroup\@href {#1}{\urlprefix }}%
\providecommand \urlprefix  [0]{URL }%
\providecommand \Eprint [0]{\href }%
\providecommand \doibase [0]{https://doi.org/}%
\providecommand \selectlanguage [0]{\@gobble}%
\providecommand \bibinfo  [0]{\@secondoftwo}%
\providecommand \bibfield  [0]{\@secondoftwo}%
\providecommand \translation [1]{[#1]}%
\providecommand \BibitemOpen [0]{}%
\providecommand \bibitemStop [0]{}%
\providecommand \bibitemNoStop [0]{.\EOS\space}%
\providecommand \EOS [0]{\spacefactor3000\relax}%
\providecommand \BibitemShut  [1]{\csname bibitem#1\endcsname}%
\let\auto@bib@innerbib\@empty
\bibitem [{\citenamefont {Josephson}(1962)}]{Josephson1962}%
  \BibitemOpen
  \bibfield  {author} {\bibinfo {author} {\bibfnamefont {B.}~\bibnamefont {Josephson}},\ }\href {https://doi.org/https://doi.org/10.1016/0031-9163(62)91369-0} {\bibfield  {journal} {\bibinfo  {journal} {Phys. Lett.}\ }\textbf {\bibinfo {volume} {1}},\ \bibinfo {pages} {251} (\bibinfo {year} {1962})}\BibitemShut {NoStop}%
\bibitem [{\citenamefont {Barone}\ and\ \citenamefont {Paterno}(1982)}]{Barone1982}%
  \BibitemOpen
  \bibfield  {author} {\bibinfo {author} {\bibfnamefont {A.}~\bibnamefont {Barone}}\ and\ \bibinfo {author} {\bibfnamefont {G.}~\bibnamefont {Paterno}},\ }\href@noop {} {\emph {\bibinfo {title} {Physics and Applications of the Josephson Effect}}}\ (\bibinfo  {publisher} {John Wiley \& Sons, Inc.},\ \bibinfo {year} {1982})\BibitemShut {NoStop}%
\bibitem [{\citenamefont {Beenakker}\ and\ \citenamefont {van Houten}(1991)}]{Beenakker1991}%
  \BibitemOpen
  \bibfield  {author} {\bibinfo {author} {\bibfnamefont {C.~W.~J.}\ \bibnamefont {Beenakker}}\ and\ \bibinfo {author} {\bibfnamefont {H.}~\bibnamefont {van Houten}},\ }\href {https://doi.org/10.1103/PhysRevLett.66.3056} {\bibfield  {journal} {\bibinfo  {journal} {Phys. Rev. Lett.}\ }\textbf {\bibinfo {volume} {66}},\ \bibinfo {pages} {3056} (\bibinfo {year} {1991})}\BibitemShut {NoStop}%
\bibitem [{\citenamefont {Furusaki}\ and\ \citenamefont {Tsukada}(1991)}]{Furusaki1991}%
  \BibitemOpen
  \bibfield  {author} {\bibinfo {author} {\bibfnamefont {A.}~\bibnamefont {Furusaki}}\ and\ \bibinfo {author} {\bibfnamefont {M.}~\bibnamefont {Tsukada}},\ }\href {https://doi.org/10.1103/PhysRevB.43.10164} {\bibfield  {journal} {\bibinfo  {journal} {Phys. Rev. B}\ }\textbf {\bibinfo {volume} {43}},\ \bibinfo {pages} {10164} (\bibinfo {year} {1991})}\BibitemShut {NoStop}%
\bibitem [{\citenamefont {Pillet}\ \emph {et~al.}(2010)\citenamefont {Pillet}, \citenamefont {Quay}, \citenamefont {Morfin}, \citenamefont {Bena}, \citenamefont {Yeyati},\ and\ \citenamefont {Joyez}}]{Pillet2010}%
  \BibitemOpen
  \bibfield  {author} {\bibinfo {author} {\bibfnamefont {J.-D.}\ \bibnamefont {Pillet}}, \bibinfo {author} {\bibfnamefont {C.~H.~L.}\ \bibnamefont {Quay}}, \bibinfo {author} {\bibfnamefont {P.}~\bibnamefont {Morfin}}, \bibinfo {author} {\bibfnamefont {C.}~\bibnamefont {Bena}}, \bibinfo {author} {\bibfnamefont {A.~L.}\ \bibnamefont {Yeyati}},\ and\ \bibinfo {author} {\bibfnamefont {P.}~\bibnamefont {Joyez}},\ }\href {https://doi.org/10.1038/nphys1811} {\bibfield  {journal} {\bibinfo  {journal} {Nat. Phys.}\ }\textbf {\bibinfo {volume} {6}},\ \bibinfo {pages} {965} (\bibinfo {year} {2010})}\BibitemShut {NoStop}%
\bibitem [{\citenamefont {Chang}\ \emph {et~al.}(2013)\citenamefont {Chang}, \citenamefont {Manucharyan}, \citenamefont {Jespersen}, \citenamefont {Nyg{\aa}rd},\ and\ \citenamefont {Marcus}}]{Chang2013}%
  \BibitemOpen
  \bibfield  {author} {\bibinfo {author} {\bibfnamefont {W.}~\bibnamefont {Chang}}, \bibinfo {author} {\bibfnamefont {V.~E.}\ \bibnamefont {Manucharyan}}, \bibinfo {author} {\bibfnamefont {T.~S.}\ \bibnamefont {Jespersen}}, \bibinfo {author} {\bibfnamefont {J.}~\bibnamefont {Nyg{\aa}rd}},\ and\ \bibinfo {author} {\bibfnamefont {C.~M.}\ \bibnamefont {Marcus}},\ }\href {https://doi.org/10.1103/PhysRevLett.110.217005} {\bibfield  {journal} {\bibinfo  {journal} {Phys. Rev. Lett.}\ }\textbf {\bibinfo {volume} {110}},\ \bibinfo {pages} {217005} (\bibinfo {year} {2013})}\BibitemShut {NoStop}%
\bibitem [{\citenamefont {Bretheau}\ \emph {et~al.}(2013{\natexlab{a}})\citenamefont {Bretheau}, \citenamefont {Girit}, \citenamefont {Pothier}, \citenamefont {Esteve},\ and\ \citenamefont {Urbina}}]{Bretheau2013}%
  \BibitemOpen
  \bibfield  {author} {\bibinfo {author} {\bibfnamefont {L.}~\bibnamefont {Bretheau}}, \bibinfo {author} {\bibfnamefont {{\c C}.~{\"O}.}\ \bibnamefont {Girit}}, \bibinfo {author} {\bibfnamefont {H.}~\bibnamefont {Pothier}}, \bibinfo {author} {\bibfnamefont {D.}~\bibnamefont {Esteve}},\ and\ \bibinfo {author} {\bibfnamefont {C.}~\bibnamefont {Urbina}},\ }\href {https://doi.org/10.1038/nature12315} {\bibfield  {journal} {\bibinfo  {journal} {Nature}\ }\textbf {\bibinfo {volume} {499}},\ \bibinfo {pages} {312} (\bibinfo {year} {2013}{\natexlab{a}})}\BibitemShut {NoStop}%
\bibitem [{\citenamefont {Bretheau}\ \emph {et~al.}(2013{\natexlab{b}})\citenamefont {Bretheau}, \citenamefont {Girit}, \citenamefont {Urbina}, \citenamefont {Esteve},\ and\ \citenamefont {Pothier}}]{Bretheau2013a}%
  \BibitemOpen
  \bibfield  {author} {\bibinfo {author} {\bibfnamefont {L.}~\bibnamefont {Bretheau}}, \bibinfo {author} {\bibfnamefont {{\c C}.~{\"O}.}\ \bibnamefont {Girit}}, \bibinfo {author} {\bibfnamefont {C.}~\bibnamefont {Urbina}}, \bibinfo {author} {\bibfnamefont {D.}~\bibnamefont {Esteve}},\ and\ \bibinfo {author} {\bibfnamefont {H.}~\bibnamefont {Pothier}},\ }\href {https://doi.org/10.1103/PhysRevX.3.041034} {\bibfield  {journal} {\bibinfo  {journal} {Phys. Rev. X}\ }\textbf {\bibinfo {volume} {3}},\ \bibinfo {pages} {041034} (\bibinfo {year} {2013}{\natexlab{b}})}\BibitemShut {NoStop}%
\bibitem [{\citenamefont {Janvier}\ \emph {et~al.}(2015)\citenamefont {Janvier}, \citenamefont {Tosi}, \citenamefont {Bretheau}, \citenamefont {Girit}, \citenamefont {Stern}, \citenamefont {Bertet}, \citenamefont {Joyez}, \citenamefont {Vion}, \citenamefont {Esteve}, \citenamefont {Goffman}, \citenamefont {Pothier},\ and\ \citenamefont {Urbina}}]{Janvier2015}%
  \BibitemOpen
  \bibfield  {author} {\bibinfo {author} {\bibfnamefont {C.}~\bibnamefont {Janvier}}, \bibinfo {author} {\bibfnamefont {L.}~\bibnamefont {Tosi}}, \bibinfo {author} {\bibfnamefont {L.}~\bibnamefont {Bretheau}}, \bibinfo {author} {\bibfnamefont {{\c C}.~{\"O}.}\ \bibnamefont {Girit}}, \bibinfo {author} {\bibfnamefont {M.}~\bibnamefont {Stern}}, \bibinfo {author} {\bibfnamefont {P.}~\bibnamefont {Bertet}}, \bibinfo {author} {\bibfnamefont {P.}~\bibnamefont {Joyez}}, \bibinfo {author} {\bibfnamefont {D.}~\bibnamefont {Vion}}, \bibinfo {author} {\bibfnamefont {D.}~\bibnamefont {Esteve}}, \bibinfo {author} {\bibfnamefont {M.~F.}\ \bibnamefont {Goffman}}, \bibinfo {author} {\bibfnamefont {H.}~\bibnamefont {Pothier}},\ and\ \bibinfo {author} {\bibfnamefont {C.}~\bibnamefont {Urbina}},\ }\href {https://doi.org/10.1126/science.aab2179} {\bibfield  {journal} {\bibinfo  {journal} {Science}\ }\textbf {\bibinfo {volume} {349}},\ \bibinfo {pages} {1199} (\bibinfo {year} {2015})}\BibitemShut {NoStop}%
\bibitem [{\citenamefont {Bretheau}\ \emph {et~al.}(2017)\citenamefont {Bretheau}, \citenamefont {Wang}, \citenamefont {Pisoni}, \citenamefont {Watanabe}, \citenamefont {Taniguchi},\ and\ \citenamefont {{Jarillo-Herrero}}}]{Bretheau2017}%
  \BibitemOpen
  \bibfield  {author} {\bibinfo {author} {\bibfnamefont {L.}~\bibnamefont {Bretheau}}, \bibinfo {author} {\bibfnamefont {J.~I.-J.}\ \bibnamefont {Wang}}, \bibinfo {author} {\bibfnamefont {R.}~\bibnamefont {Pisoni}}, \bibinfo {author} {\bibfnamefont {K.}~\bibnamefont {Watanabe}}, \bibinfo {author} {\bibfnamefont {T.}~\bibnamefont {Taniguchi}},\ and\ \bibinfo {author} {\bibfnamefont {P.}~\bibnamefont {{Jarillo-Herrero}}},\ }\href {https://doi.org/10.1038/nphys4110} {\bibfield  {journal} {\bibinfo  {journal} {Nat. Phys.}\ }\textbf {\bibinfo {volume} {13}},\ \bibinfo {pages} {756} (\bibinfo {year} {2017})}\BibitemShut {NoStop}%
\bibitem [{\citenamefont {{van Woerkom}}\ \emph {et~al.}(2017)\citenamefont {{van Woerkom}}, \citenamefont {Proutski}, \citenamefont {{van Heck}}, \citenamefont {Bouman}, \citenamefont {V{\"a}yrynen}, \citenamefont {Glazman}, \citenamefont {Krogstrup}, \citenamefont {Nyg{\aa}rd}, \citenamefont {Kouwenhoven},\ and\ \citenamefont {Geresdi}}]{vanWoerkom2017}%
  \BibitemOpen
  \bibfield  {author} {\bibinfo {author} {\bibfnamefont {D.~J.}\ \bibnamefont {{van Woerkom}}}, \bibinfo {author} {\bibfnamefont {A.}~\bibnamefont {Proutski}}, \bibinfo {author} {\bibfnamefont {B.}~\bibnamefont {{van Heck}}}, \bibinfo {author} {\bibfnamefont {D.}~\bibnamefont {Bouman}}, \bibinfo {author} {\bibfnamefont {J.~I.}\ \bibnamefont {V{\"a}yrynen}}, \bibinfo {author} {\bibfnamefont {L.~I.}\ \bibnamefont {Glazman}}, \bibinfo {author} {\bibfnamefont {P.}~\bibnamefont {Krogstrup}}, \bibinfo {author} {\bibfnamefont {J.}~\bibnamefont {Nyg{\aa}rd}}, \bibinfo {author} {\bibfnamefont {L.~P.}\ \bibnamefont {Kouwenhoven}},\ and\ \bibinfo {author} {\bibfnamefont {A.}~\bibnamefont {Geresdi}},\ }\href {https://doi.org/10.1038/nphys4150} {\bibfield  {journal} {\bibinfo  {journal} {Nat. Phys.}\ }\textbf {\bibinfo {volume} {13}},\ \bibinfo {pages} {876} (\bibinfo {year} {2017})}\BibitemShut {NoStop}%
\bibitem [{\citenamefont {Hays}\ \emph {et~al.}(2018)\citenamefont {Hays}, \citenamefont {{de Lange}}, \citenamefont {Serniak}, \citenamefont {{van Woerkom}}, \citenamefont {Bouman}, \citenamefont {Krogstrup}, \citenamefont {Nyg{\aa}rd}, \citenamefont {Geresdi},\ and\ \citenamefont {Devoret}}]{Hays2018}%
  \BibitemOpen
  \bibfield  {author} {\bibinfo {author} {\bibfnamefont {M.}~\bibnamefont {Hays}}, \bibinfo {author} {\bibfnamefont {G.}~\bibnamefont {{de Lange}}}, \bibinfo {author} {\bibfnamefont {K.}~\bibnamefont {Serniak}}, \bibinfo {author} {\bibfnamefont {D.~J.}\ \bibnamefont {{van Woerkom}}}, \bibinfo {author} {\bibfnamefont {D.}~\bibnamefont {Bouman}}, \bibinfo {author} {\bibfnamefont {P.}~\bibnamefont {Krogstrup}}, \bibinfo {author} {\bibfnamefont {J.}~\bibnamefont {Nyg{\aa}rd}}, \bibinfo {author} {\bibfnamefont {A.}~\bibnamefont {Geresdi}},\ and\ \bibinfo {author} {\bibfnamefont {M.~H.}\ \bibnamefont {Devoret}},\ }\href {https://doi.org/10.1103/PhysRevLett.121.047001} {\bibfield  {journal} {\bibinfo  {journal} {Phys. Rev. Lett.}\ }\textbf {\bibinfo {volume} {121}},\ \bibinfo {pages} {047001} (\bibinfo {year} {2018})}\BibitemShut {NoStop}%
\bibitem [{\citenamefont {Tosi}\ \emph {et~al.}(2019)\citenamefont {Tosi}, \citenamefont {Metzger}, \citenamefont {Goffman}, \citenamefont {Urbina}, \citenamefont {Pothier}, \citenamefont {Park}, \citenamefont {Yeyati}, \citenamefont {Nyg{\aa}rd},\ and\ \citenamefont {Krogstrup}}]{Tosi2019}%
  \BibitemOpen
  \bibfield  {author} {\bibinfo {author} {\bibfnamefont {L.}~\bibnamefont {Tosi}}, \bibinfo {author} {\bibfnamefont {C.}~\bibnamefont {Metzger}}, \bibinfo {author} {\bibfnamefont {M.~F.}\ \bibnamefont {Goffman}}, \bibinfo {author} {\bibfnamefont {C.}~\bibnamefont {Urbina}}, \bibinfo {author} {\bibfnamefont {H.}~\bibnamefont {Pothier}}, \bibinfo {author} {\bibfnamefont {S.}~\bibnamefont {Park}}, \bibinfo {author} {\bibfnamefont {A.~L.}\ \bibnamefont {Yeyati}}, \bibinfo {author} {\bibfnamefont {J.}~\bibnamefont {Nyg{\aa}rd}},\ and\ \bibinfo {author} {\bibfnamefont {P.}~\bibnamefont {Krogstrup}},\ }\href {https://doi.org/10.1103/PhysRevX.9.011010} {\bibfield  {journal} {\bibinfo  {journal} {Phys. Rev. X}\ }\textbf {\bibinfo {volume} {9}},\ \bibinfo {pages} {011010} (\bibinfo {year} {2019})}\BibitemShut {NoStop}%
\bibitem [{\citenamefont {Nichele}\ \emph {et~al.}(2020)\citenamefont {Nichele}, \citenamefont {Portol{\'e}s}, \citenamefont {Fornieri}, \citenamefont {Whiticar}, \citenamefont {Drachmann}, \citenamefont {Gronin}, \citenamefont {Wang}, \citenamefont {Gardner}, \citenamefont {Thomas}, \citenamefont {Hatke}, \citenamefont {Manfra},\ and\ \citenamefont {Marcus}}]{Nichele2020}%
  \BibitemOpen
  \bibfield  {author} {\bibinfo {author} {\bibfnamefont {F.}~\bibnamefont {Nichele}}, \bibinfo {author} {\bibfnamefont {E.}~\bibnamefont {Portol{\'e}s}}, \bibinfo {author} {\bibfnamefont {A.}~\bibnamefont {Fornieri}}, \bibinfo {author} {\bibfnamefont {A.~M.}\ \bibnamefont {Whiticar}}, \bibinfo {author} {\bibfnamefont {A.~C.~C.}\ \bibnamefont {Drachmann}}, \bibinfo {author} {\bibfnamefont {S.}~\bibnamefont {Gronin}}, \bibinfo {author} {\bibfnamefont {T.}~\bibnamefont {Wang}}, \bibinfo {author} {\bibfnamefont {G.~C.}\ \bibnamefont {Gardner}}, \bibinfo {author} {\bibfnamefont {C.}~\bibnamefont {Thomas}}, \bibinfo {author} {\bibfnamefont {A.~T.}\ \bibnamefont {Hatke}}, \bibinfo {author} {\bibfnamefont {M.~J.}\ \bibnamefont {Manfra}},\ and\ \bibinfo {author} {\bibfnamefont {C.~M.}\ \bibnamefont {Marcus}},\ }\href {https://doi.org/10.1103/PhysRevLett.124.226801} {\bibfield  {journal} {\bibinfo  {journal} {Phys. Rev. Lett.}\ }\textbf {\bibinfo {volume} {124}},\ \bibinfo {pages} {226801} (\bibinfo {year}
  {2020})}\BibitemShut {NoStop}%
\bibitem [{\citenamefont {Hays}\ \emph {et~al.}(2020)\citenamefont {Hays}, \citenamefont {Fatemi}, \citenamefont {Serniak}, \citenamefont {Bouman}, \citenamefont {Diamond}, \citenamefont {de~Lange}, \citenamefont {Krogstrup}, \citenamefont {Nyg{\aa}rd}, \citenamefont {Geresdi},\ and\ \citenamefont {Devoret}}]{Hays2020}%
  \BibitemOpen
  \bibfield  {author} {\bibinfo {author} {\bibfnamefont {M.}~\bibnamefont {Hays}}, \bibinfo {author} {\bibfnamefont {V.}~\bibnamefont {Fatemi}}, \bibinfo {author} {\bibfnamefont {K.}~\bibnamefont {Serniak}}, \bibinfo {author} {\bibfnamefont {D.}~\bibnamefont {Bouman}}, \bibinfo {author} {\bibfnamefont {S.}~\bibnamefont {Diamond}}, \bibinfo {author} {\bibfnamefont {G.}~\bibnamefont {de~Lange}}, \bibinfo {author} {\bibfnamefont {P.}~\bibnamefont {Krogstrup}}, \bibinfo {author} {\bibfnamefont {J.}~\bibnamefont {Nyg{\aa}rd}}, \bibinfo {author} {\bibfnamefont {A.}~\bibnamefont {Geresdi}},\ and\ \bibinfo {author} {\bibfnamefont {M.~H.}\ \bibnamefont {Devoret}},\ }\href {https://doi.org/10.1038/s41567-020-0952-3} {\bibfield  {journal} {\bibinfo  {journal} {Nat. Phys.}\ }\textbf {\bibinfo {volume} {16}},\ \bibinfo {pages} {1103} (\bibinfo {year} {2020})}\BibitemShut {NoStop}%
\bibitem [{\citenamefont {Hays}\ \emph {et~al.}(2021)\citenamefont {Hays}, \citenamefont {Fatemi}, \citenamefont {Bouman}, \citenamefont {Cerrillo}, \citenamefont {Diamond}, \citenamefont {Serniak}, \citenamefont {Connolly}, \citenamefont {Krogstrup}, \citenamefont {Nyg{\aa}rd}, \citenamefont {Levy~Yeyati}, \citenamefont {Geresdi},\ and\ \citenamefont {Devoret}}]{Hays2021}%
  \BibitemOpen
  \bibfield  {author} {\bibinfo {author} {\bibfnamefont {M.}~\bibnamefont {Hays}}, \bibinfo {author} {\bibfnamefont {V.}~\bibnamefont {Fatemi}}, \bibinfo {author} {\bibfnamefont {D.}~\bibnamefont {Bouman}}, \bibinfo {author} {\bibfnamefont {J.}~\bibnamefont {Cerrillo}}, \bibinfo {author} {\bibfnamefont {S.}~\bibnamefont {Diamond}}, \bibinfo {author} {\bibfnamefont {K.}~\bibnamefont {Serniak}}, \bibinfo {author} {\bibfnamefont {T.}~\bibnamefont {Connolly}}, \bibinfo {author} {\bibfnamefont {P.}~\bibnamefont {Krogstrup}}, \bibinfo {author} {\bibfnamefont {J.}~\bibnamefont {Nyg{\aa}rd}}, \bibinfo {author} {\bibfnamefont {A.}~\bibnamefont {Levy~Yeyati}}, \bibinfo {author} {\bibfnamefont {A.}~\bibnamefont {Geresdi}},\ and\ \bibinfo {author} {\bibfnamefont {M.~H.}\ \bibnamefont {Devoret}},\ }\href {https://doi.org/10.1126/science.abf0345} {\bibfield  {journal} {\bibinfo  {journal} {Science}\ }\textbf {\bibinfo {volume} {373}},\ \bibinfo {pages} {430} (\bibinfo {year} {2021})}\BibitemShut {NoStop}%
\bibitem [{\citenamefont {{Pita-Vidal}}\ \emph {et~al.}(2023)\citenamefont {{Pita-Vidal}}, \citenamefont {Bargerbos}, \citenamefont {{\v Z}itko}, \citenamefont {Splitthoff}, \citenamefont {Gr{\"u}nhaupt}, \citenamefont {Wesdorp}, \citenamefont {Liu}, \citenamefont {Kouwenhoven}, \citenamefont {Aguado}, \citenamefont {{van Heck}}, \citenamefont {Kou},\ and\ \citenamefont {Andersen}}]{Pita-Vidal2023}%
  \BibitemOpen
  \bibfield  {author} {\bibinfo {author} {\bibfnamefont {M.}~\bibnamefont {{Pita-Vidal}}}, \bibinfo {author} {\bibfnamefont {A.}~\bibnamefont {Bargerbos}}, \bibinfo {author} {\bibfnamefont {R.}~\bibnamefont {{\v Z}itko}}, \bibinfo {author} {\bibfnamefont {L.~J.}\ \bibnamefont {Splitthoff}}, \bibinfo {author} {\bibfnamefont {L.}~\bibnamefont {Gr{\"u}nhaupt}}, \bibinfo {author} {\bibfnamefont {J.~J.}\ \bibnamefont {Wesdorp}}, \bibinfo {author} {\bibfnamefont {Y.}~\bibnamefont {Liu}}, \bibinfo {author} {\bibfnamefont {L.~P.}\ \bibnamefont {Kouwenhoven}}, \bibinfo {author} {\bibfnamefont {R.}~\bibnamefont {Aguado}}, \bibinfo {author} {\bibfnamefont {B.}~\bibnamefont {{van Heck}}}, \bibinfo {author} {\bibfnamefont {A.}~\bibnamefont {Kou}},\ and\ \bibinfo {author} {\bibfnamefont {C.~K.}\ \bibnamefont {Andersen}},\ }\href {https://doi.org/10.1038/s41567-023-02071-x} {\bibfield  {journal} {\bibinfo  {journal} {Nat. Phys.}\ }\textbf {\bibinfo {volume} {19}},\ \bibinfo {pages} {1110} (\bibinfo {year}
  {2023})}\BibitemShut {NoStop}%
\bibitem [{\citenamefont {Hinderling}\ \emph {et~al.}(2023)\citenamefont {Hinderling}, \citenamefont {Sabonis}, \citenamefont {Paredes}, \citenamefont {Haxell}, \citenamefont {Coraiola}, \citenamefont {ten Kate}, \citenamefont {Cheah}, \citenamefont {Krizek}, \citenamefont {Schott}, \citenamefont {Wegscheider},\ and\ \citenamefont {Nichele}}]{Hinderling2023}%
  \BibitemOpen
  \bibfield  {author} {\bibinfo {author} {\bibfnamefont {M.}~\bibnamefont {Hinderling}}, \bibinfo {author} {\bibfnamefont {D.}~\bibnamefont {Sabonis}}, \bibinfo {author} {\bibfnamefont {S.}~\bibnamefont {Paredes}}, \bibinfo {author} {\bibfnamefont {D.~Z.}\ \bibnamefont {Haxell}}, \bibinfo {author} {\bibfnamefont {M.}~\bibnamefont {Coraiola}}, \bibinfo {author} {\bibfnamefont {S.~C.}\ \bibnamefont {ten Kate}}, \bibinfo {author} {\bibfnamefont {E.}~\bibnamefont {Cheah}}, \bibinfo {author} {\bibfnamefont {F.}~\bibnamefont {Krizek}}, \bibinfo {author} {\bibfnamefont {R.}~\bibnamefont {Schott}}, \bibinfo {author} {\bibfnamefont {W.}~\bibnamefont {Wegscheider}},\ and\ \bibinfo {author} {\bibfnamefont {F.}~\bibnamefont {Nichele}},\ }\href {https://doi.org/10.1103/PhysRevApplied.19.054026} {\bibfield  {journal} {\bibinfo  {journal} {Phys. Rev. Appl.}\ }\textbf {\bibinfo {volume} {19}},\ \bibinfo {pages} {054026} (\bibinfo {year} {2023})}\BibitemShut {NoStop}%
\bibitem [{\citenamefont {Wesdorp}\ \emph {et~al.}(2023)\citenamefont {Wesdorp}, \citenamefont {Matute-Caňadas}, \citenamefont {Vaartjes}, \citenamefont {Grünhaupt}, \citenamefont {Laeven}, \citenamefont {Roelofs}, \citenamefont {Splitthoff}, \citenamefont {Pita-Vidal}, \citenamefont {Bargerbos}, \citenamefont {van Woerkom}, \citenamefont {Krogstrup}, \citenamefont {Kouwenhoven}, \citenamefont {Andersen}, \citenamefont {Yeyati}, \citenamefont {van Heck},\ and\ \citenamefont {de~Lange}}]{Wesdorp2022}%
  \BibitemOpen
  \bibfield  {author} {\bibinfo {author} {\bibfnamefont {J.~J.}\ \bibnamefont {Wesdorp}}, \bibinfo {author} {\bibfnamefont {F.~J.}\ \bibnamefont {Matute-Caňadas}}, \bibinfo {author} {\bibfnamefont {A.}~\bibnamefont {Vaartjes}}, \bibinfo {author} {\bibfnamefont {L.}~\bibnamefont {Grünhaupt}}, \bibinfo {author} {\bibfnamefont {T.}~\bibnamefont {Laeven}}, \bibinfo {author} {\bibfnamefont {S.}~\bibnamefont {Roelofs}}, \bibinfo {author} {\bibfnamefont {L.~J.}\ \bibnamefont {Splitthoff}}, \bibinfo {author} {\bibfnamefont {M.}~\bibnamefont {Pita-Vidal}}, \bibinfo {author} {\bibfnamefont {A.}~\bibnamefont {Bargerbos}}, \bibinfo {author} {\bibfnamefont {D.~J.}\ \bibnamefont {van Woerkom}}, \bibinfo {author} {\bibfnamefont {P.}~\bibnamefont {Krogstrup}}, \bibinfo {author} {\bibfnamefont {L.~P.}\ \bibnamefont {Kouwenhoven}}, \bibinfo {author} {\bibfnamefont {C.~K.}\ \bibnamefont {Andersen}}, \bibinfo {author} {\bibfnamefont {A.~L.}\ \bibnamefont {Yeyati}}, \bibinfo {author} {\bibfnamefont {B.}~\bibnamefont {van
  Heck}},\ and\ \bibinfo {author} {\bibfnamefont {G.}~\bibnamefont {de~Lange}},\ }\href {https://journals.aps.org/prb/accepted/08070KefM6a1d70d36999a05dcb79f3d9ab4286ed} {\bibfield  {journal} {\bibinfo  {journal} {Phys. Rev. B}\ } (\bibinfo {year} {2023})}\BibitemShut {NoStop}%
\bibitem [{\citenamefont {Draelos}\ \emph {et~al.}(2019)\citenamefont {Draelos}, \citenamefont {Wei}, \citenamefont {Seredinski}, \citenamefont {Li}, \citenamefont {Mehta}, \citenamefont {Watanabe}, \citenamefont {Taniguchi}, \citenamefont {Borzenets}, \citenamefont {Amet},\ and\ \citenamefont {Finkelstein}}]{Draelos2019}%
  \BibitemOpen
  \bibfield  {author} {\bibinfo {author} {\bibfnamefont {A.~W.}\ \bibnamefont {Draelos}}, \bibinfo {author} {\bibfnamefont {M.-T.}\ \bibnamefont {Wei}}, \bibinfo {author} {\bibfnamefont {A.}~\bibnamefont {Seredinski}}, \bibinfo {author} {\bibfnamefont {H.}~\bibnamefont {Li}}, \bibinfo {author} {\bibfnamefont {Y.}~\bibnamefont {Mehta}}, \bibinfo {author} {\bibfnamefont {K.}~\bibnamefont {Watanabe}}, \bibinfo {author} {\bibfnamefont {T.}~\bibnamefont {Taniguchi}}, \bibinfo {author} {\bibfnamefont {I.~V.}\ \bibnamefont {Borzenets}}, \bibinfo {author} {\bibfnamefont {F.}~\bibnamefont {Amet}},\ and\ \bibinfo {author} {\bibfnamefont {G.}~\bibnamefont {Finkelstein}},\ }\href {https://doi.org/10.1021/acs.nanolett.8b04330} {\bibfield  {journal} {\bibinfo  {journal} {Nano Lett.}\ }\textbf {\bibinfo {volume} {19}},\ \bibinfo {pages} {1039} (\bibinfo {year} {2019})}\BibitemShut {NoStop}%
\bibitem [{\citenamefont {Graziano}\ \emph {et~al.}(2020)\citenamefont {Graziano}, \citenamefont {Lee}, \citenamefont {Pendharkar}, \citenamefont {Palmstr{\o}m},\ and\ \citenamefont {Pribiag}}]{Graziano2020}%
  \BibitemOpen
  \bibfield  {author} {\bibinfo {author} {\bibfnamefont {G.~V.}\ \bibnamefont {Graziano}}, \bibinfo {author} {\bibfnamefont {J.~S.}\ \bibnamefont {Lee}}, \bibinfo {author} {\bibfnamefont {M.}~\bibnamefont {Pendharkar}}, \bibinfo {author} {\bibfnamefont {C.~J.}\ \bibnamefont {Palmstr{\o}m}},\ and\ \bibinfo {author} {\bibfnamefont {V.~S.}\ \bibnamefont {Pribiag}},\ }\href {https://doi.org/10.1103/PhysRevB.101.054510} {\bibfield  {journal} {\bibinfo  {journal} {Phys. Rev. B}\ }\textbf {\bibinfo {volume} {101}},\ \bibinfo {pages} {054510} (\bibinfo {year} {2020})}\BibitemShut {NoStop}%
\bibitem [{\citenamefont {Pankratova}\ \emph {et~al.}(2020)\citenamefont {Pankratova}, \citenamefont {Lee}, \citenamefont {Kuzmin}, \citenamefont {Wickramasinghe}, \citenamefont {Mayer}, \citenamefont {Yuan}, \citenamefont {Vavilov}, \citenamefont {Shabani},\ and\ \citenamefont {Manucharyan}}]{Pankratova2020}%
  \BibitemOpen
  \bibfield  {author} {\bibinfo {author} {\bibfnamefont {N.}~\bibnamefont {Pankratova}}, \bibinfo {author} {\bibfnamefont {H.}~\bibnamefont {Lee}}, \bibinfo {author} {\bibfnamefont {R.}~\bibnamefont {Kuzmin}}, \bibinfo {author} {\bibfnamefont {K.}~\bibnamefont {Wickramasinghe}}, \bibinfo {author} {\bibfnamefont {W.}~\bibnamefont {Mayer}}, \bibinfo {author} {\bibfnamefont {J.}~\bibnamefont {Yuan}}, \bibinfo {author} {\bibfnamefont {M.~G.}\ \bibnamefont {Vavilov}}, \bibinfo {author} {\bibfnamefont {J.}~\bibnamefont {Shabani}},\ and\ \bibinfo {author} {\bibfnamefont {V.~E.}\ \bibnamefont {Manucharyan}},\ }\href {https://doi.org/10.1103/PhysRevX.10.031051} {\bibfield  {journal} {\bibinfo  {journal} {Phys. Rev. X}\ }\textbf {\bibinfo {volume} {10}},\ \bibinfo {pages} {031051} (\bibinfo {year} {2020})}\BibitemShut {NoStop}%
\bibitem [{\citenamefont {Arnault}\ \emph {et~al.}(2021)\citenamefont {Arnault}, \citenamefont {Larson}, \citenamefont {Seredinski}, \citenamefont {Zhao}, \citenamefont {Idris}, \citenamefont {McConnell}, \citenamefont {Watanabe}, \citenamefont {Taniguchi}, \citenamefont {Borzenets}, \citenamefont {Amet},\ and\ \citenamefont {Finkelstein}}]{Arnault2021}%
  \BibitemOpen
  \bibfield  {author} {\bibinfo {author} {\bibfnamefont {E.~G.}\ \bibnamefont {Arnault}}, \bibinfo {author} {\bibfnamefont {T.~F.~Q.}\ \bibnamefont {Larson}}, \bibinfo {author} {\bibfnamefont {A.}~\bibnamefont {Seredinski}}, \bibinfo {author} {\bibfnamefont {L.}~\bibnamefont {Zhao}}, \bibinfo {author} {\bibfnamefont {S.}~\bibnamefont {Idris}}, \bibinfo {author} {\bibfnamefont {A.}~\bibnamefont {McConnell}}, \bibinfo {author} {\bibfnamefont {K.}~\bibnamefont {Watanabe}}, \bibinfo {author} {\bibfnamefont {T.}~\bibnamefont {Taniguchi}}, \bibinfo {author} {\bibfnamefont {I.}~\bibnamefont {Borzenets}}, \bibinfo {author} {\bibfnamefont {F.}~\bibnamefont {Amet}},\ and\ \bibinfo {author} {\bibfnamefont {G.}~\bibnamefont {Finkelstein}},\ }\href {https://doi.org/10.1021/acs.nanolett.1c03474} {\bibfield  {journal} {\bibinfo  {journal} {Nano Lett.}\ }\textbf {\bibinfo {volume} {21}},\ \bibinfo {pages} {9668} (\bibinfo {year} {2021})}\BibitemShut {NoStop}%
\bibitem [{\citenamefont {Graziano}\ \emph {et~al.}(2022)\citenamefont {Graziano}, \citenamefont {Gupta}, \citenamefont {Pendharkar}, \citenamefont {Dong}, \citenamefont {Dempsey}, \citenamefont {Palmstr{\o}m},\ and\ \citenamefont {Pribiag}}]{Graziano2022}%
  \BibitemOpen
  \bibfield  {author} {\bibinfo {author} {\bibfnamefont {G.~V.}\ \bibnamefont {Graziano}}, \bibinfo {author} {\bibfnamefont {M.}~\bibnamefont {Gupta}}, \bibinfo {author} {\bibfnamefont {M.}~\bibnamefont {Pendharkar}}, \bibinfo {author} {\bibfnamefont {J.~T.}\ \bibnamefont {Dong}}, \bibinfo {author} {\bibfnamefont {C.~P.}\ \bibnamefont {Dempsey}}, \bibinfo {author} {\bibfnamefont {C.}~\bibnamefont {Palmstr{\o}m}},\ and\ \bibinfo {author} {\bibfnamefont {V.~S.}\ \bibnamefont {Pribiag}},\ }\href@noop {} {\bibfield  {journal} {\bibinfo  {journal} {Nat. Commun.}\ }\textbf {\bibinfo {volume} {13}},\ \bibinfo {pages} {5933} (\bibinfo {year} {2022})}\BibitemShut {NoStop}%
\bibitem [{\citenamefont {Gupta}\ \emph {et~al.}(2023)\citenamefont {Gupta}, \citenamefont {Graziano}, \citenamefont {Pendharkar}, \citenamefont {Dong}, \citenamefont {Dempsey}, \citenamefont {Palmstr{\o}m},\ and\ \citenamefont {Pribiag}}]{Gupta2023}%
  \BibitemOpen
  \bibfield  {author} {\bibinfo {author} {\bibfnamefont {M.}~\bibnamefont {Gupta}}, \bibinfo {author} {\bibfnamefont {G.~V.}\ \bibnamefont {Graziano}}, \bibinfo {author} {\bibfnamefont {M.}~\bibnamefont {Pendharkar}}, \bibinfo {author} {\bibfnamefont {J.~T.}\ \bibnamefont {Dong}}, \bibinfo {author} {\bibfnamefont {C.~P.}\ \bibnamefont {Dempsey}}, \bibinfo {author} {\bibfnamefont {C.}~\bibnamefont {Palmstr{\o}m}},\ and\ \bibinfo {author} {\bibfnamefont {V.~S.}\ \bibnamefont {Pribiag}},\ }\href@noop {} {\bibfield  {journal} {\bibinfo  {journal} {Nat. Commun.}\ }\textbf {\bibinfo {volume} {14}},\ \bibinfo {pages} {3078} (\bibinfo {year} {2023})}\BibitemShut {NoStop}%
\bibitem [{\citenamefont {Yokoyama}\ and\ \citenamefont {Nazarov}(2015)}]{Yokoyama2015}%
  \BibitemOpen
  \bibfield  {author} {\bibinfo {author} {\bibfnamefont {T.}~\bibnamefont {Yokoyama}}\ and\ \bibinfo {author} {\bibfnamefont {Y.~V.}\ \bibnamefont {Nazarov}},\ }\href {https://doi.org/10.1103/PhysRevB.92.155437} {\bibfield  {journal} {\bibinfo  {journal} {Phys. Rev. B}\ }\textbf {\bibinfo {volume} {92}},\ \bibinfo {pages} {155437} (\bibinfo {year} {2015})}\BibitemShut {NoStop}%
\bibitem [{\citenamefont {Riwar}\ \emph {et~al.}(2016)\citenamefont {Riwar}, \citenamefont {Houzet}, \citenamefont {Meyer},\ and\ \citenamefont {Nazarov}}]{Riwar2016}%
  \BibitemOpen
  \bibfield  {author} {\bibinfo {author} {\bibfnamefont {R.-P.}\ \bibnamefont {Riwar}}, \bibinfo {author} {\bibfnamefont {M.}~\bibnamefont {Houzet}}, \bibinfo {author} {\bibfnamefont {J.~S.}\ \bibnamefont {Meyer}},\ and\ \bibinfo {author} {\bibfnamefont {Y.~V.}\ \bibnamefont {Nazarov}},\ }\href {https://doi.org/10.1038/ncomms11167} {\bibfield  {journal} {\bibinfo  {journal} {Nat. Commun.}\ }\textbf {\bibinfo {volume} {7}},\ \bibinfo {pages} {11167} (\bibinfo {year} {2016})}\BibitemShut {NoStop}%
\bibitem [{\citenamefont {Eriksson}\ \emph {et~al.}(2017)\citenamefont {Eriksson}, \citenamefont {Riwar}, \citenamefont {Houzet}, \citenamefont {Meyer},\ and\ \citenamefont {Nazarov}}]{Eriksson2017}%
  \BibitemOpen
  \bibfield  {author} {\bibinfo {author} {\bibfnamefont {E.}~\bibnamefont {Eriksson}}, \bibinfo {author} {\bibfnamefont {R.-P.}\ \bibnamefont {Riwar}}, \bibinfo {author} {\bibfnamefont {M.}~\bibnamefont {Houzet}}, \bibinfo {author} {\bibfnamefont {J.~S.}\ \bibnamefont {Meyer}},\ and\ \bibinfo {author} {\bibfnamefont {Y.~V.}\ \bibnamefont {Nazarov}},\ }\href {https://doi.org/10.1103/PhysRevB.95.075417} {\bibfield  {journal} {\bibinfo  {journal} {Phys. Rev. B}\ }\textbf {\bibinfo {volume} {95}},\ \bibinfo {pages} {075417} (\bibinfo {year} {2017})}\BibitemShut {NoStop}%
\bibitem [{\citenamefont {Meyer}\ and\ \citenamefont {Houzet}(2017)}]{Meyer2017}%
  \BibitemOpen
  \bibfield  {author} {\bibinfo {author} {\bibfnamefont {J.~S.}\ \bibnamefont {Meyer}}\ and\ \bibinfo {author} {\bibfnamefont {M.}~\bibnamefont {Houzet}},\ }\href {https://doi.org/10.1103/PhysRevLett.119.136807} {\bibfield  {journal} {\bibinfo  {journal} {Phys. Rev. Lett.}\ }\textbf {\bibinfo {volume} {119}},\ \bibinfo {pages} {136807} (\bibinfo {year} {2017})}\BibitemShut {NoStop}%
\bibitem [{\citenamefont {Xie}\ \emph {et~al.}(2017)\citenamefont {Xie}, \citenamefont {Vavilov},\ and\ \citenamefont {Levchenko}}]{Xie2017}%
  \BibitemOpen
  \bibfield  {author} {\bibinfo {author} {\bibfnamefont {H.-Y.}\ \bibnamefont {Xie}}, \bibinfo {author} {\bibfnamefont {M.~G.}\ \bibnamefont {Vavilov}},\ and\ \bibinfo {author} {\bibfnamefont {A.}~\bibnamefont {Levchenko}},\ }\href {https://doi.org/10.1103/PhysRevB.96.161406} {\bibfield  {journal} {\bibinfo  {journal} {Phys. Rev. B}\ }\textbf {\bibinfo {volume} {96}},\ \bibinfo {pages} {161406(R)} (\bibinfo {year} {2017})}\BibitemShut {NoStop}%
\bibitem [{\citenamefont {Xie}\ and\ \citenamefont {Levchenko}(2019)}]{Xie2019}%
  \BibitemOpen
  \bibfield  {author} {\bibinfo {author} {\bibfnamefont {H.-Y.}\ \bibnamefont {Xie}}\ and\ \bibinfo {author} {\bibfnamefont {A.}~\bibnamefont {Levchenko}},\ }\href {https://doi.org/10.1103/PhysRevB.99.094519} {\bibfield  {journal} {\bibinfo  {journal} {Phys. Rev. B}\ }\textbf {\bibinfo {volume} {99}},\ \bibinfo {pages} {094519} (\bibinfo {year} {2019})}\BibitemShut {NoStop}%
\bibitem [{\citenamefont {Repin}\ \emph {et~al.}(2019)\citenamefont {Repin}, \citenamefont {Chen},\ and\ \citenamefont {Nazarov}}]{Repin2019}%
  \BibitemOpen
  \bibfield  {author} {\bibinfo {author} {\bibfnamefont {E.~V.}\ \bibnamefont {Repin}}, \bibinfo {author} {\bibfnamefont {Y.}~\bibnamefont {Chen}},\ and\ \bibinfo {author} {\bibfnamefont {Y.~V.}\ \bibnamefont {Nazarov}},\ }\href {https://doi.org/10.1103/PhysRevB.99.165414} {\bibfield  {journal} {\bibinfo  {journal} {Phys. Rev. B}\ }\textbf {\bibinfo {volume} {99}},\ \bibinfo {pages} {165414} (\bibinfo {year} {2019})}\BibitemShut {NoStop}%
\bibitem [{\citenamefont {Peralta~Gavensky}\ \emph {et~al.}(2019)\citenamefont {Peralta~Gavensky}, \citenamefont {Usaj},\ and\ \citenamefont {Balseiro}}]{PeraltaGavensky2019}%
  \BibitemOpen
  \bibfield  {author} {\bibinfo {author} {\bibfnamefont {L.}~\bibnamefont {Peralta~Gavensky}}, \bibinfo {author} {\bibfnamefont {G.}~\bibnamefont {Usaj}},\ and\ \bibinfo {author} {\bibfnamefont {C.~A.}\ \bibnamefont {Balseiro}},\ }\href {https://doi.org/10.1103/PhysRevB.100.014514} {\bibfield  {journal} {\bibinfo  {journal} {Phys. Rev. B}\ }\textbf {\bibinfo {volume} {100}},\ \bibinfo {pages} {014514} (\bibinfo {year} {2019})}\BibitemShut {NoStop}%
\bibitem [{\citenamefont {Houzet}\ and\ \citenamefont {Meyer}(2019)}]{Houzet2019}%
  \BibitemOpen
  \bibfield  {author} {\bibinfo {author} {\bibfnamefont {M.}~\bibnamefont {Houzet}}\ and\ \bibinfo {author} {\bibfnamefont {J.~S.}\ \bibnamefont {Meyer}},\ }\href {https://doi.org/10.1103/PhysRevB.100.014521} {\bibfield  {journal} {\bibinfo  {journal} {Phys. Rev. B}\ }\textbf {\bibinfo {volume} {100}},\ \bibinfo {pages} {014521} (\bibinfo {year} {2019})}\BibitemShut {NoStop}%
\bibitem [{\citenamefont {Klees}\ \emph {et~al.}(2020)\citenamefont {Klees}, \citenamefont {Rastelli}, \citenamefont {Cuevas},\ and\ \citenamefont {Belzig}}]{Klees2020}%
  \BibitemOpen
  \bibfield  {author} {\bibinfo {author} {\bibfnamefont {R.~L.}\ \bibnamefont {Klees}}, \bibinfo {author} {\bibfnamefont {G.}~\bibnamefont {Rastelli}}, \bibinfo {author} {\bibfnamefont {J.~C.}\ \bibnamefont {Cuevas}},\ and\ \bibinfo {author} {\bibfnamefont {W.}~\bibnamefont {Belzig}},\ }\href {https://doi.org/10.1103/PhysRevLett.124.197002} {\bibfield  {journal} {\bibinfo  {journal} {Phys. Rev. Lett.}\ }\textbf {\bibinfo {volume} {124}},\ \bibinfo {pages} {197002} (\bibinfo {year} {2020})}\BibitemShut {NoStop}%
\bibitem [{\citenamefont {Weisbrich}\ \emph {et~al.}(2021)\citenamefont {Weisbrich}, \citenamefont {Klees}, \citenamefont {Rastelli},\ and\ \citenamefont {Belzig}}]{Weisbrich2021}%
  \BibitemOpen
  \bibfield  {author} {\bibinfo {author} {\bibfnamefont {H.}~\bibnamefont {Weisbrich}}, \bibinfo {author} {\bibfnamefont {R.~L.}\ \bibnamefont {Klees}}, \bibinfo {author} {\bibfnamefont {G.}~\bibnamefont {Rastelli}},\ and\ \bibinfo {author} {\bibfnamefont {W.}~\bibnamefont {Belzig}},\ }\href {https://doi.org/10.1103/PRXQuantum.2.010310} {\bibfield  {journal} {\bibinfo  {journal} {PRX Quantum}\ }\textbf {\bibinfo {volume} {2}},\ \bibinfo {pages} {010310} (\bibinfo {year} {2021})}\BibitemShut {NoStop}%
\bibitem [{\citenamefont {Xie}\ \emph {et~al.}(2022)\citenamefont {Xie}, \citenamefont {Hasan},\ and\ \citenamefont {Levchenko}}]{Xie2022}%
  \BibitemOpen
  \bibfield  {author} {\bibinfo {author} {\bibfnamefont {H.-Y.}\ \bibnamefont {Xie}}, \bibinfo {author} {\bibfnamefont {J.}~\bibnamefont {Hasan}},\ and\ \bibinfo {author} {\bibfnamefont {A.}~\bibnamefont {Levchenko}},\ }\href {https://doi.org/10.1103/PhysRevB.105.L241404} {\bibfield  {journal} {\bibinfo  {journal} {Phys. Rev. B}\ }\textbf {\bibinfo {volume} {105}},\ \bibinfo {pages} {L241404} (\bibinfo {year} {2022})}\BibitemShut {NoStop}%
\bibitem [{\citenamefont {Barakov}\ and\ \citenamefont {Nazarov}(2023)}]{Barakov2023}%
  \BibitemOpen
  \bibfield  {author} {\bibinfo {author} {\bibfnamefont {H.}~\bibnamefont {Barakov}}\ and\ \bibinfo {author} {\bibfnamefont {Y.~V.}\ \bibnamefont {Nazarov}},\ }\href {https://doi.org/10.1103/PhysRevB.107.014507} {\bibfield  {journal} {\bibinfo  {journal} {Phys. Rev. B}\ }\textbf {\bibinfo {volume} {107}},\ \bibinfo {pages} {014507} (\bibinfo {year} {2023})}\BibitemShut {NoStop}%
\bibitem [{\citenamefont {Teshler}\ \emph {et~al.}(2023)\citenamefont {Teshler}, \citenamefont {Weisbrich}, \citenamefont {Sturm}, \citenamefont {Klees}, \citenamefont {Rastelli},\ and\ \citenamefont {Belzig}}]{Teshler2023}%
  \BibitemOpen
  \bibfield  {author} {\bibinfo {author} {\bibfnamefont {L.}~\bibnamefont {Teshler}}, \bibinfo {author} {\bibfnamefont {H.}~\bibnamefont {Weisbrich}}, \bibinfo {author} {\bibfnamefont {J.}~\bibnamefont {Sturm}}, \bibinfo {author} {\bibfnamefont {R.~L.}\ \bibnamefont {Klees}}, \bibinfo {author} {\bibfnamefont {G.}~\bibnamefont {Rastelli}},\ and\ \bibinfo {author} {\bibfnamefont {W.}~\bibnamefont {Belzig}},\ }\href {https://doi.org/10.21468/SciPostPhys.15.5.214} {\bibfield  {journal} {\bibinfo  {journal} {SciPost Phys.}\ }\textbf {\bibinfo {volume} {15}},\ \bibinfo {pages} {214} (\bibinfo {year} {2023})}\BibitemShut {NoStop}%
\bibitem [{\citenamefont {Coraiola}\ \emph {et~al.}(2023)\citenamefont {Coraiola}, \citenamefont {Haxell}, \citenamefont {Sabonis}, \citenamefont {Weisbrich}, \citenamefont {Svetogorov}, \citenamefont {Hinderling}, \citenamefont {{ten Kate}}, \citenamefont {Cheah}, \citenamefont {Krizek}, \citenamefont {Schott}, \citenamefont {Wegscheider}, \citenamefont {Cuevas}, \citenamefont {Belzig},\ and\ \citenamefont {Nichele}}]{Coraiola2023}%
  \BibitemOpen
  \bibfield  {author} {\bibinfo {author} {\bibfnamefont {M.}~\bibnamefont {Coraiola}}, \bibinfo {author} {\bibfnamefont {D.~Z.}\ \bibnamefont {Haxell}}, \bibinfo {author} {\bibfnamefont {D.}~\bibnamefont {Sabonis}}, \bibinfo {author} {\bibfnamefont {H.}~\bibnamefont {Weisbrich}}, \bibinfo {author} {\bibfnamefont {A.~E.}\ \bibnamefont {Svetogorov}}, \bibinfo {author} {\bibfnamefont {M.}~\bibnamefont {Hinderling}}, \bibinfo {author} {\bibfnamefont {S.~C.}\ \bibnamefont {{ten Kate}}}, \bibinfo {author} {\bibfnamefont {E.}~\bibnamefont {Cheah}}, \bibinfo {author} {\bibfnamefont {F.}~\bibnamefont {Krizek}}, \bibinfo {author} {\bibfnamefont {R.}~\bibnamefont {Schott}}, \bibinfo {author} {\bibfnamefont {W.}~\bibnamefont {Wegscheider}}, \bibinfo {author} {\bibfnamefont {J.~C.}\ \bibnamefont {Cuevas}}, \bibinfo {author} {\bibfnamefont {W.}~\bibnamefont {Belzig}},\ and\ \bibinfo {author} {\bibfnamefont {F.}~\bibnamefont {Nichele}},\ }\href {https://doi.org/10.1038/s41467-023-42356-6} {\bibfield  {journal} {\bibinfo
  {journal} {Nat. Commun.}\ }\textbf {\bibinfo {volume} {14}},\ \bibinfo {pages} {6784} (\bibinfo {year} {2023})}\BibitemShut {NoStop}%
\bibitem [{\citenamefont {Antonelli}\ \emph {et~al.}(2025)\citenamefont {Antonelli}, \citenamefont {Coraiola}, \citenamefont {Ohnmacht}, \citenamefont {Svetogorov}, \citenamefont {Sabonis}, \citenamefont {ten Kate}, \citenamefont {Cheah}, \citenamefont {Krizek}, \citenamefont {Schott}, \citenamefont {Cuevas}, \citenamefont {Belzig}, \citenamefont {Wegscheider},\ and\ \citenamefont {Nichele}}]{Antonelli2025}%
  \BibitemOpen
  \bibfield  {author} {\bibinfo {author} {\bibfnamefont {T.}~\bibnamefont {Antonelli}}, \bibinfo {author} {\bibfnamefont {M.}~\bibnamefont {Coraiola}}, \bibinfo {author} {\bibfnamefont {D.~C.}\ \bibnamefont {Ohnmacht}}, \bibinfo {author} {\bibfnamefont {A.~E.}\ \bibnamefont {Svetogorov}}, \bibinfo {author} {\bibfnamefont {D.}~\bibnamefont {Sabonis}}, \bibinfo {author} {\bibfnamefont {S.~C.}\ \bibnamefont {ten Kate}}, \bibinfo {author} {\bibfnamefont {E.}~\bibnamefont {Cheah}}, \bibinfo {author} {\bibfnamefont {F.}~\bibnamefont {Krizek}}, \bibinfo {author} {\bibfnamefont {R.}~\bibnamefont {Schott}}, \bibinfo {author} {\bibfnamefont {J.~C.}\ \bibnamefont {Cuevas}}, \bibinfo {author} {\bibfnamefont {W.}~\bibnamefont {Belzig}}, \bibinfo {author} {\bibfnamefont {W.}~\bibnamefont {Wegscheider}},\ and\ \bibinfo {author} {\bibfnamefont {F.}~\bibnamefont {Nichele}},\ }\href {https://arxiv.org/abs/2501.07982} {\bibinfo {title} {Exploring the energy spectrum of a four-terminal josephson junction: Towards topological
  andreev band structures}} (\bibinfo {year} {2025}),\ \Eprint {https://arxiv.org/abs/2501.07982} {arXiv:2501.07982 [cond-mat.mes-hall]} \BibitemShut {NoStop}%
\bibitem [{\citenamefont {Coraiola}\ \emph {et~al.}(2024{\natexlab{a}})\citenamefont {Coraiola}, \citenamefont {Svetogorov}, \citenamefont {Haxell}, \citenamefont {Sabonis}, \citenamefont {Hinderling}, \citenamefont {Ten~Kate}, \citenamefont {Cheah}, \citenamefont {Krizek}, \citenamefont {Schott}, \citenamefont {Wegscheider} \emph {et~al.}}]{Coraiola2024}%
  \BibitemOpen
  \bibfield  {author} {\bibinfo {author} {\bibfnamefont {M.}~\bibnamefont {Coraiola}}, \bibinfo {author} {\bibfnamefont {A.~E.}\ \bibnamefont {Svetogorov}}, \bibinfo {author} {\bibfnamefont {D.~Z.}\ \bibnamefont {Haxell}}, \bibinfo {author} {\bibfnamefont {D.}~\bibnamefont {Sabonis}}, \bibinfo {author} {\bibfnamefont {M.}~\bibnamefont {Hinderling}}, \bibinfo {author} {\bibfnamefont {S.~C.}\ \bibnamefont {Ten~Kate}}, \bibinfo {author} {\bibfnamefont {E.}~\bibnamefont {Cheah}}, \bibinfo {author} {\bibfnamefont {F.}~\bibnamefont {Krizek}}, \bibinfo {author} {\bibfnamefont {R.}~\bibnamefont {Schott}}, \bibinfo {author} {\bibfnamefont {W.}~\bibnamefont {Wegscheider}}, \emph {et~al.},\ }\href@noop {} {\bibfield  {journal} {\bibinfo  {journal} {ACS Nano}\ }\textbf {\bibinfo {volume} {18}},\ \bibinfo {pages} {9221} (\bibinfo {year} {2024}{\natexlab{a}})}\BibitemShut {NoStop}%
\bibitem [{\citenamefont {Ohnmacht}\ \emph {et~al.}(2025{\natexlab{a}})\citenamefont {Ohnmacht}, \citenamefont {Wilhelm}, \citenamefont {Weisbrich},\ and\ \citenamefont {Belzig}}]{Ohnmacht2025}%
  \BibitemOpen
  \bibfield  {author} {\bibinfo {author} {\bibfnamefont {D.~C.}\ \bibnamefont {Ohnmacht}}, \bibinfo {author} {\bibfnamefont {V.}~\bibnamefont {Wilhelm}}, \bibinfo {author} {\bibfnamefont {H.}~\bibnamefont {Weisbrich}},\ and\ \bibinfo {author} {\bibfnamefont {W.}~\bibnamefont {Belzig}},\ }\href@noop {} {\bibfield  {journal} {\bibinfo  {journal} {Phys. Rev. Lett.}\ }\textbf {\bibinfo {volume} {134}},\ \bibinfo {pages} {156601} (\bibinfo {year} {2025}{\natexlab{a}})}\BibitemShut {NoStop}%
\bibitem [{\citenamefont {Freyn}\ \emph {et~al.}(2011)\citenamefont {Freyn}, \citenamefont {Dou{\c c}ot}, \citenamefont {Feinberg},\ and\ \citenamefont {M{\'e}lin}}]{Freyn2011}%
  \BibitemOpen
  \bibfield  {author} {\bibinfo {author} {\bibfnamefont {A.}~\bibnamefont {Freyn}}, \bibinfo {author} {\bibfnamefont {B.}~\bibnamefont {Dou{\c c}ot}}, \bibinfo {author} {\bibfnamefont {D.}~\bibnamefont {Feinberg}},\ and\ \bibinfo {author} {\bibfnamefont {R.}~\bibnamefont {M{\'e}lin}},\ }\href {https://doi.org/10.1103/PhysRevLett.106.257005} {\bibfield  {journal} {\bibinfo  {journal} {Phys. Rev. Lett.}\ }\textbf {\bibinfo {volume} {106}},\ \bibinfo {pages} {257005} (\bibinfo {year} {2011})}\BibitemShut {NoStop}%
\bibitem [{\citenamefont {Jonckheere}\ \emph {et~al.}(2013)\citenamefont {Jonckheere}, \citenamefont {Rech}, \citenamefont {Martin}, \citenamefont {Dou{\c c}ot}, \citenamefont {Feinberg},\ and\ \citenamefont {M{\'e}lin}}]{Jonckheere2013}%
  \BibitemOpen
  \bibfield  {author} {\bibinfo {author} {\bibfnamefont {T.}~\bibnamefont {Jonckheere}}, \bibinfo {author} {\bibfnamefont {J.}~\bibnamefont {Rech}}, \bibinfo {author} {\bibfnamefont {T.}~\bibnamefont {Martin}}, \bibinfo {author} {\bibfnamefont {B.}~\bibnamefont {Dou{\c c}ot}}, \bibinfo {author} {\bibfnamefont {D.}~\bibnamefont {Feinberg}},\ and\ \bibinfo {author} {\bibfnamefont {R.}~\bibnamefont {M{\'e}lin}},\ }\href {https://doi.org/10.1103/PhysRevB.87.214501} {\bibfield  {journal} {\bibinfo  {journal} {Phys. Rev. B}\ }\textbf {\bibinfo {volume} {87}},\ \bibinfo {pages} {214501} (\bibinfo {year} {2013})}\BibitemShut {NoStop}%
\bibitem [{\citenamefont {Rech}\ \emph {et~al.}(2014)\citenamefont {Rech}, \citenamefont {Jonckheere}, \citenamefont {Martin}, \citenamefont {Dou{\c c}ot}, \citenamefont {Feinberg},\ and\ \citenamefont {M{\'e}lin}}]{Rech2014}%
  \BibitemOpen
  \bibfield  {author} {\bibinfo {author} {\bibfnamefont {J.}~\bibnamefont {Rech}}, \bibinfo {author} {\bibfnamefont {T.}~\bibnamefont {Jonckheere}}, \bibinfo {author} {\bibfnamefont {T.}~\bibnamefont {Martin}}, \bibinfo {author} {\bibfnamefont {B.}~\bibnamefont {Dou{\c c}ot}}, \bibinfo {author} {\bibfnamefont {D.}~\bibnamefont {Feinberg}},\ and\ \bibinfo {author} {\bibfnamefont {R.}~\bibnamefont {M{\'e}lin}},\ }\href {https://doi.org/10.1103/PhysRevB.90.075419} {\bibfield  {journal} {\bibinfo  {journal} {Phys. Rev. B}\ }\textbf {\bibinfo {volume} {90}},\ \bibinfo {pages} {075419} (\bibinfo {year} {2014})}\BibitemShut {NoStop}%
\bibitem [{\citenamefont {Feinberg}\ \emph {et~al.}(2015)\citenamefont {Feinberg}, \citenamefont {Jonckheere}, \citenamefont {Rech}, \citenamefont {Martin}, \citenamefont {Dou{\c c}ot},\ and\ \citenamefont {M{\'e}lin}}]{Feinberg2015}%
  \BibitemOpen
  \bibfield  {author} {\bibinfo {author} {\bibfnamefont {D.}~\bibnamefont {Feinberg}}, \bibinfo {author} {\bibfnamefont {T.}~\bibnamefont {Jonckheere}}, \bibinfo {author} {\bibfnamefont {J.}~\bibnamefont {Rech}}, \bibinfo {author} {\bibfnamefont {T.}~\bibnamefont {Martin}}, \bibinfo {author} {\bibfnamefont {B.}~\bibnamefont {Dou{\c c}ot}},\ and\ \bibinfo {author} {\bibfnamefont {R.}~\bibnamefont {M{\'e}lin}},\ }\href {https://doi.org/10.1140/epjb/e2015-50849-3} {\bibfield  {journal} {\bibinfo  {journal} {Eur. Phys. J. B}\ }\textbf {\bibinfo {volume} {88}},\ \bibinfo {pages} {99} (\bibinfo {year} {2015})}\BibitemShut {NoStop}%
\bibitem [{\citenamefont {Melo}\ \emph {et~al.}(2022)\citenamefont {Melo}, \citenamefont {Fatemi},\ and\ \citenamefont {Akhmerov}}]{Melo2022}%
  \BibitemOpen
  \bibfield  {author} {\bibinfo {author} {\bibfnamefont {A.}~\bibnamefont {Melo}}, \bibinfo {author} {\bibfnamefont {V.}~\bibnamefont {Fatemi}},\ and\ \bibinfo {author} {\bibfnamefont {A.~R.}\ \bibnamefont {Akhmerov}},\ }\href {https://doi.org/10.21468/SciPostPhys.12.1.017} {\bibfield  {journal} {\bibinfo  {journal} {SciPost Phys.}\ }\textbf {\bibinfo {volume} {12}},\ \bibinfo {pages} {017} (\bibinfo {year} {2022})}\BibitemShut {NoStop}%
\bibitem [{\citenamefont {M{\'e}lin}\ \emph {et~al.}(2023)\citenamefont {M{\'e}lin}, \citenamefont {Danneau},\ and\ \citenamefont {Winkelmann}}]{Melin2023a}%
  \BibitemOpen
  \bibfield  {author} {\bibinfo {author} {\bibfnamefont {R.}~\bibnamefont {M{\'e}lin}}, \bibinfo {author} {\bibfnamefont {R.}~\bibnamefont {Danneau}},\ and\ \bibinfo {author} {\bibfnamefont {C.~B.}\ \bibnamefont {Winkelmann}},\ }\href {https://doi.org/10.1103/PhysRevResearch.5.033124} {\bibfield  {journal} {\bibinfo  {journal} {Phys. Rev. Res.}\ }\textbf {\bibinfo {volume} {5}},\ \bibinfo {pages} {033124} (\bibinfo {year} {2023})}\BibitemShut {NoStop}%
\bibitem [{\citenamefont {Pfeffer}\ \emph {et~al.}(2014)\citenamefont {Pfeffer}, \citenamefont {Duvauchelle}, \citenamefont {Courtois}, \citenamefont {M{\'e}lin}, \citenamefont {Feinberg},\ and\ \citenamefont {Lefloch}}]{Pfeffer2014}%
  \BibitemOpen
  \bibfield  {author} {\bibinfo {author} {\bibfnamefont {A.~H.}\ \bibnamefont {Pfeffer}}, \bibinfo {author} {\bibfnamefont {J.~E.}\ \bibnamefont {Duvauchelle}}, \bibinfo {author} {\bibfnamefont {H.}~\bibnamefont {Courtois}}, \bibinfo {author} {\bibfnamefont {R.}~\bibnamefont {M{\'e}lin}}, \bibinfo {author} {\bibfnamefont {D.}~\bibnamefont {Feinberg}},\ and\ \bibinfo {author} {\bibfnamefont {F.}~\bibnamefont {Lefloch}},\ }\href {https://doi.org/10.1103/PhysRevB.90.075401} {\bibfield  {journal} {\bibinfo  {journal} {Phys. Rev. B}\ }\textbf {\bibinfo {volume} {90}},\ \bibinfo {pages} {075401} (\bibinfo {year} {2014})}\BibitemShut {NoStop}%
\bibitem [{\citenamefont {Cohen}\ \emph {et~al.}(2018)\citenamefont {Cohen}, \citenamefont {Ronen}, \citenamefont {Kang}, \citenamefont {Heiblum}, \citenamefont {Feinberg}, \citenamefont {M{\'e}lin},\ and\ \citenamefont {Shtrikman}}]{Cohen2018}%
  \BibitemOpen
  \bibfield  {author} {\bibinfo {author} {\bibfnamefont {Y.}~\bibnamefont {Cohen}}, \bibinfo {author} {\bibfnamefont {Y.}~\bibnamefont {Ronen}}, \bibinfo {author} {\bibfnamefont {J.-H.}\ \bibnamefont {Kang}}, \bibinfo {author} {\bibfnamefont {M.}~\bibnamefont {Heiblum}}, \bibinfo {author} {\bibfnamefont {D.}~\bibnamefont {Feinberg}}, \bibinfo {author} {\bibfnamefont {R.}~\bibnamefont {M{\'e}lin}},\ and\ \bibinfo {author} {\bibfnamefont {H.}~\bibnamefont {Shtrikman}},\ }\href {https://doi.org/10.1073/pnas.1800044115} {\bibfield  {journal} {\bibinfo  {journal} {Proc. Natl. Acad. Sci. U.S.A.}\ }\textbf {\bibinfo {volume} {115}},\ \bibinfo {pages} {6991} (\bibinfo {year} {2018})}\BibitemShut {NoStop}%
\bibitem [{\citenamefont {Huang}\ \emph {et~al.}(2022)\citenamefont {Huang}, \citenamefont {Ronen}, \citenamefont {M{\'e}lin}, \citenamefont {Feinberg}, \citenamefont {Watanabe}, \citenamefont {Taniguchi},\ and\ \citenamefont {Kim}}]{Huang2022}%
  \BibitemOpen
  \bibfield  {author} {\bibinfo {author} {\bibfnamefont {K.-F.}\ \bibnamefont {Huang}}, \bibinfo {author} {\bibfnamefont {Y.}~\bibnamefont {Ronen}}, \bibinfo {author} {\bibfnamefont {R.}~\bibnamefont {M{\'e}lin}}, \bibinfo {author} {\bibfnamefont {D.}~\bibnamefont {Feinberg}}, \bibinfo {author} {\bibfnamefont {K.}~\bibnamefont {Watanabe}}, \bibinfo {author} {\bibfnamefont {T.}~\bibnamefont {Taniguchi}},\ and\ \bibinfo {author} {\bibfnamefont {P.}~\bibnamefont {Kim}},\ }\href {https://doi.org/10.1038/s41467-022-30732-7} {\bibfield  {journal} {\bibinfo  {journal} {Nat. Commun.}\ }\textbf {\bibinfo {volume} {13}},\ \bibinfo {pages} {3032} (\bibinfo {year} {2022})}\BibitemShut {NoStop}%
\bibitem [{\citenamefont {Arnault}\ \emph {et~al.}(2022)\citenamefont {Arnault}, \citenamefont {Idris}, \citenamefont {McConnell}, \citenamefont {Zhao}, \citenamefont {Larson}, \citenamefont {Watanabe}, \citenamefont {Taniguchi}, \citenamefont {Finkelstein},\ and\ \citenamefont {Amet}}]{Arnault2022}%
  \BibitemOpen
  \bibfield  {author} {\bibinfo {author} {\bibfnamefont {E.~G.}\ \bibnamefont {Arnault}}, \bibinfo {author} {\bibfnamefont {S.}~\bibnamefont {Idris}}, \bibinfo {author} {\bibfnamefont {A.}~\bibnamefont {McConnell}}, \bibinfo {author} {\bibfnamefont {L.}~\bibnamefont {Zhao}}, \bibinfo {author} {\bibfnamefont {T.~F.}\ \bibnamefont {Larson}}, \bibinfo {author} {\bibfnamefont {K.}~\bibnamefont {Watanabe}}, \bibinfo {author} {\bibfnamefont {T.}~\bibnamefont {Taniguchi}}, \bibinfo {author} {\bibfnamefont {G.}~\bibnamefont {Finkelstein}},\ and\ \bibinfo {author} {\bibfnamefont {F.}~\bibnamefont {Amet}},\ }\href {https://doi.org/10.1021/acs.nanolett.2c01999} {\bibfield  {journal} {\bibinfo  {journal} {Nano Lett.}\ }\textbf {\bibinfo {volume} {22}},\ \bibinfo {pages} {7073} (\bibinfo {year} {2022})}\BibitemShut {NoStop}%
\bibitem [{\citenamefont {Ohnmacht}\ \emph {et~al.}(2024)\citenamefont {Ohnmacht}, \citenamefont {Coraiola}, \citenamefont {Garc\'{\i}a-Esteban}, \citenamefont {Sabonis}, \citenamefont {Nichele}, \citenamefont {Belzig},\ and\ \citenamefont {Cuevas}}]{Ohnmacht2024}%
  \BibitemOpen
  \bibfield  {author} {\bibinfo {author} {\bibfnamefont {D.~C.}\ \bibnamefont {Ohnmacht}}, \bibinfo {author} {\bibfnamefont {M.}~\bibnamefont {Coraiola}}, \bibinfo {author} {\bibfnamefont {J.~J.}\ \bibnamefont {Garc\'{\i}a-Esteban}}, \bibinfo {author} {\bibfnamefont {D.}~\bibnamefont {Sabonis}}, \bibinfo {author} {\bibfnamefont {F.}~\bibnamefont {Nichele}}, \bibinfo {author} {\bibfnamefont {W.}~\bibnamefont {Belzig}},\ and\ \bibinfo {author} {\bibfnamefont {J.~C.}\ \bibnamefont {Cuevas}},\ }\href {https://doi.org/10.1103/PhysRevB.109.L241407} {\bibfield  {journal} {\bibinfo  {journal} {Phys. Rev. B}\ }\textbf {\bibinfo {volume} {109}},\ \bibinfo {pages} {L241407} (\bibinfo {year} {2024})}\BibitemShut {NoStop}%
\bibitem [{\citenamefont {Ohnmacht}\ \emph {et~al.}(2025{\natexlab{b}})\citenamefont {Ohnmacht}, \citenamefont {Wilhelm},\ and\ \citenamefont {Belzig}}]{Ohnmacht2025a}%
  \BibitemOpen
  \bibfield  {author} {\bibinfo {author} {\bibfnamefont {D.~C.}\ \bibnamefont {Ohnmacht}}, \bibinfo {author} {\bibfnamefont {V.}~\bibnamefont {Wilhelm}},\ and\ \bibinfo {author} {\bibfnamefont {W.}~\bibnamefont {Belzig}},\ }\href {https://doi.org/10.48550/arXiv.2503.10874} {\bibinfo {title} {Reflectionless modes as a source of {{Weyl}} nodes in multiterminal {{Josephson}} junctions}} (\bibinfo {year} {2025}{\natexlab{b}}),\ \Eprint {https://arxiv.org/abs/2503.10874} {arXiv:2503.10874 [cond-mat]} \BibitemShut {NoStop}%
\bibitem [{\citenamefont {Coraiola}\ \emph {et~al.}(2024{\natexlab{b}})\citenamefont {Coraiola}, \citenamefont {Haxell}, \citenamefont {Sabonis}, \citenamefont {Hinderling}, \citenamefont {Kate}, \citenamefont {Cheah}, \citenamefont {Krizek}, \citenamefont {Schott}, \citenamefont {Wegscheider},\ and\ \citenamefont {Nichele}}]{Coraiola2023a}%
  \BibitemOpen
  \bibfield  {author} {\bibinfo {author} {\bibfnamefont {M.}~\bibnamefont {Coraiola}}, \bibinfo {author} {\bibfnamefont {D.~Z.}\ \bibnamefont {Haxell}}, \bibinfo {author} {\bibfnamefont {D.}~\bibnamefont {Sabonis}}, \bibinfo {author} {\bibfnamefont {M.}~\bibnamefont {Hinderling}}, \bibinfo {author} {\bibfnamefont {S.~C.~t.}\ \bibnamefont {Kate}}, \bibinfo {author} {\bibfnamefont {E.}~\bibnamefont {Cheah}}, \bibinfo {author} {\bibfnamefont {F.}~\bibnamefont {Krizek}}, \bibinfo {author} {\bibfnamefont {R.}~\bibnamefont {Schott}}, \bibinfo {author} {\bibfnamefont {W.}~\bibnamefont {Wegscheider}},\ and\ \bibinfo {author} {\bibfnamefont {F.}~\bibnamefont {Nichele}},\ }\href {https://doi.org/10.1103/PhysRevX.14.031024} {\bibfield  {journal} {\bibinfo  {journal} {Phys. Rev. X}\ }\textbf {\bibinfo {volume} {14}},\ \bibinfo {pages} {031024} (\bibinfo {year} {2024}{\natexlab{b}})}\BibitemShut {NoStop}%
\end{thebibliography}%

\end{document}